\documentclass[sigconf]{acmart}

\usepackage{tabularx} 
\usepackage{multirow}
\usepackage{framed,acmart-taps}
\usepackage{xcolor}

\definecolor{shadecolor}{gray}{0.9} 

\newenvironment{acmbox}[1]{%
    \MakeFramed {\advance\hsize-\width \FrameRestore}%
    \noindent\textbf{#1}\par\smallskip
}{%
    \endMakeFramed
}

\usepackage{array}
\usepackage{graphicx}    
\usepackage{booktabs}    

\usepackage{xcolor} %
\newif\ifshowrevs
\showrevstrue  

\ifshowrevs
  
\else
  
\fi

\AtBeginDocument{%
  }

\copyrightyear{2026}
\acmYear{2026}
\setcopyright{cc}
\setcctype{by}
\acmConference[CHI '26]{Proceedings of the 2026 CHI Conference on Human Factors in Computing Systems}{April 13--17, 2026}{Barcelona, Spain}
\acmBooktitle{Proceedings of the 2026 CHI Conference on Human Factors in Computing Systems (CHI '26), April 13--17, 2026, Barcelona, Spain}
\acmDOI{10.1145/3772318.3790724}
\acmISBN{979-8-4007-2278-3/2026/04}

\begin{document}

\title{How Do We Evaluate Experiences in Immersive Environments?}

\author{Xiang Li}
\orcid{0000-0001-5529-071X}
\affiliation{%
  \institution{University of Cambridge}
  \city{Cambridge}
  \country{United Kingdom}
}
\email{xl529@cam.ac.uk}

\author{Wei He}
\orcid{0009-0001-6745-2752}
\affiliation{%
  \institution{The Hong Kong University of Science and Technology (Guangzhou)}
  \city{Guangzhou}
  \country{China}
}
\email{whe694@connect.hkust-gz.edu.cn}

\author{Per Ola Kristensson}
\orcid{0000-0002-7139-871X}
\affiliation{%
\institution{University of Cambridge}
\city{Cambridge} 
\country{United Kingdom} 
}
\email{pok21@cam.ac.uk}

\renewcommand{\shortauthors}{Xiang Li, Wei He, and Per Ola Kristensson}

\begin{abstract}

How do we evaluate experiences in immersive environments? Despite decades of research in immersive technologies such as virtual reality, the field remains fragmented. Studies rely on overlapping constructs, heterogeneous instruments, and little agreement on what counts as immersive experience. To better understand this landscape, we conducted a bottom-up scoping review of 375 papers published in ACM CHI, UIST, VRST, SUI, IEEE VR, ISMAR, and TVCG. Our analysis reveals that evaluation practices are often domain- and purpose-specific, shaped more by local choices than by shared standards. Yet this diversity also points to new directions. Instead of multiplying instruments, researchers benefit from integrating and refining them into smarter measures. Rather than focusing only on system outputs, evaluations must center the user’s lived experience. Computational modeling offers opportunities to bridge signals across methods, but lasting progress requires open and sustainable evaluation practices that support comparability and reuse. Ultimately, our contribution is to map current practices and outline a forward-looking agenda for immersive experience research.

\end{abstract}




\maketitle

\section{Introduction}

How do we evaluate experiences in immersive environments? Despite decades of research, the field continues to struggle with fragmented definitions, inconsistent methods, and limited comparability across studies. Across the Human-Computer Interaction (HCI) community, the term ``immersive experience'' typically spans a wide range of interactive systems, including virtual reality (VR), augmented and mixed reality (AR/MR), 360-degree media, immersive games, simulation and training environments, and collaborative or productivity tools. Although these technologies vary in display properties and interaction modalities, the shared ambition is to create situations in which users feel perceptually, cognitively, or socially engrossed in a mediated environment. While prior work and textbooks have articulated foundational concepts and evaluation principles, less is known about how immersive experience is \emph{actually evaluated in practice} across technologies and domains.

Immersion has often been described as the extent to which users feel absorbed in a digital environment across many different media and technologies, including messages, games, narratives, and extended reality (XR)~\cite{IxDF2023Immersion,chen2024being}, but the term itself remains elusive and shifts across contexts and devices~\cite{slater_2003,skarbez_2017}. Crucially, the constructs used to describe immersive experience do not apply uniformly across technologies. Presence has a long history in VR research~\cite{xiao2025concept}, where it refers to place and plausibility illusions, while AR and MR systems often foreground spatial alignment, situational awareness, and task-integrated usability rather than a strict sense of ``being there.'' Similarly, embodiment has been decomposed into agency, self-location, and body ownership~\cite{Kilteni_2012,li2025bend,li2025evaluating}, yet it remains difficult to capture consistently outside controlled VR scenarios. These conceptual differences do not weaken the relevance of immersive experience; rather, they matter because they shape what gets measured, how measures are combined, and how findings transfer—or fail to transfer—across technologies and domains.

\textbf{Why does measurement matter?} In application areas central to the HCI and XR community, such as gaming and entertainment, learning and training, health and rehabilitation, perception and psychophysics, and collaborative or productivity systems, the effectiveness of an immersive system depends on how users experience it. Engagement, comfort, spatial understanding, social connection, and emotional response all shape whether a system achieves its intended purpose. Consequently, the ability to measure immersive experience reliably is not merely a theoretical concern but a practical requirement for building, validating, and comparing interactive systems.

Over time, researchers have assembled an expanding toolbox of evaluation strategies. Questionnaires dominate, especially in presence research~\cite{souza_2021}, but are frequently modified, inconsistently cited, or combined in ad hoc ways. Task and system performance measures are central in navigation and spatial cognition~\cite{nash_2000,bowman_2007}; physiological and neurophysiological signals are increasingly proposed as objective complements to self-report~\cite{dey_2020,zhang_2020}; multimodal approaches that integrate behavioral traces with subjective accounts are gaining traction~\cite{robert_2024}. Yet adding instruments does not automatically yield better evaluation. Administering many scales risks redundancy, participant fatigue, and diluted meaning. What remains unclear is not simply which measures exist, but how they are selected, combined, and deployed in practice across domains.

\textbf{What does this paper do?} To address these challenges, we conducted a bottom-up scoping review of immersive experience evaluation as it is practiced in the HCI and XR literature. Our goal is not to impose a universal definition of immersive experience, nor to prescribe evaluation guidelines, but to provide an empirical snapshot of the field by documenting how immersive experience has been defined and measured across technologies and domains, and how constructs, measures, and design goals interact in practice.

From an initial corpus of 650 papers, we coded 375 publications drawn from IEEE Transactions on Visualization and Computer Graphics (\textbf{IEEE TVCG})\footnote{\url{https://www.computer.org/csdl/journal/tg}}, IEEE Conference on Virtual Reality and 3D User Interfaces (\textbf{IEEE VR})\footnote{\url{https://ieeevr.org}}, IEEE International Symposium on Mixed and Augmented Reality (\textbf{IEEE ISMAR})\footnote{\url{https://www.ismar.net}}, ACM Conference on Human Factors in Computing Systems (\textbf{ACM CHI})\footnote{\url{https://dl.acm.org/conference/chi}}, ACM Symposium on User Interface Software and Technology (\textbf{ACM UIST})\footnote{\url{https://dl.acm.org/conference/uist}}, ACM Symposium on Virtual Reality Software and Technology (\textbf{ACM VRST})\footnote{\url{https://dl.acm.org/conference/vrst}}, and ACM Symposium on Spatial User Interaction (\textbf{ACM SUI})\footnote{\url{https://dl.acm.org/conference/sui}}. We mapped how immersive experience has been conceptualized and assessed across three decades, producing a fine-grained account of constructs, measures, and contexts. This synthesis highlights recurring patterns, exposes inconsistencies, and identifies gaps that are difficult to discern from individual studies or high-level methodological treatments.

Our synthesis yields four cross-cutting reflections. First, evaluation is domain-sensitive. For example, gaming and entertainment foreground engagement and enjoyment; health and rehabilitation emphasize safety and symptom reduction; locomotion prioritizes task efficiency and simulator sickness; perception and sensory work relies on psychophysics and fidelity. These differences are rooted in domain priorities rather than arbitrary choice. Second, research practice remains largely researcher-centered. Constructs convenient to measure often displace what users value, while qualitative and user-centered approaches remain underrepresented despite their ability to surface empathy, social connectedness, and affective resonance. Third, computational modeling offers opportunities to formalize and predict aspects of experience, from fatigue in mid-air gestures~\cite{yi2022nicer} to ergonomics in VR workspaces~\cite{fischer2024sim2vr}, provided models are empirically grounded and aligned with behavioral and self-report evidence. Fourth, open infrastructures are scarce. Limited sharing of protocols, datasets, and validated instruments constrains comparability and reproducibility, preventing cumulative progress.

\textbf{Why does this matter?} The immediate problem practitioners face is not only conceptual disagreement but \emph{measurement opacity}. Researchers and designers frequently encounter uncertainty about which instrument to use, when to deploy it, and how to implement it well. Instruments are adapted without clear justification or validation, scales are combined without an organizing rationale, and reporting practices often omit essential details about timing, ordering, training, and analysis. By consolidating how measures are actually used across domains, our review functions as an evidence-based orientation to current practice. Its contribution lies not in prescribing best practices, but in making prevailing practices visible and comparable, enabling more transparent and reflective methodological choices.

\textbf{What this paper does not do.} We do not propose a new definition of immersive experience, nor do we attempt to resolve terminological debates. Our contribution is to consolidate practices, make assumptions explicit, and outline directions for coordinated progress.

\textbf{Who is this paper for?} The paper serves newcomers seeking an empirically grounded entry point, XR and HCI researchers situating their methodological choices, and designers and practitioners who require insight into how immersive experience has been evaluated across contexts. More broadly, it invites the community to treat evaluation as a shared methodological challenge that benefits from openness, reflexivity, and collaboration.

\section{Related Work}

Existing work has made substantial contributions to clarifying key constructs of immersive experience, developing measures to capture them, and surveying evaluation practices in XR. Several foundational textbooks, for example, the Handbook of Virtual Environments\cite{hale2014handbook}, The VR Book\cite{jerald_human-centered_2017}, and 3D User Interfaces\cite{laviola20173d}, have provided broader syntheses of immersive system design and evaluation that cut across constructs, methods, and technologies. These volumes underscore both the richness of the field and the continued difficulty of organizing its diverse perspectives into a coherent whole. Yet, these efforts remain fragmented. Most prior reviews focus on a single construct, such as presence~\cite{souza_2021,skarbez_2017,felton_2022}, embodiment~\cite{Kilteni_2012,peck_2024}, or empathy~\cite{lee_2024}; a single method, such as questionnaires~\cite{souza_2021} or physiological measures~\cite{dey_2020}; or a single technology, such as AR~\cite{dey_2018,zhou_2008,kim_2018} or VR~\cite{kim_2020}. While these studies provide valuable insights, what is still missing is a comprehensive synthesis that systematically classifies how immersive experience has been conceptualized and measured across different domains, devices, and decades. 

\subsection{Foundations of Immersive Experience Constructs}
The study of immersive experience has long been entangled with debates about terminology. Researchers have struggled to disentangle immersion, as the objective properties of a system, from presence, as the user’s subjective response to those properties. Classic papers have proposed more precise distinctions, such as differentiating immersion from involvement and emotional response~\cite{slater_2003}, or framing presence as a perceptual illusion grounded in sensorimotor contingencies and plausibility~\cite{slater_2009,slater_2010}. Other works introduced taxonomies of presence, highlighting its multiple forms and determinants~\cite{slater_1994,nash_2000,sanchez_2005}. More recently, surveys have attempted to unify competing definitions, arguing for the adoption of models such as Place Illusion and Plausibility Illusion to bring conceptual clarity~\cite{skarbez_2017,felton_2022}. Despite these efforts, presence remains a contested concept, with some even questioning whether it should be conflated with higher-order constructs like engagement or absorption~\cite{Murphy_2020}. As XR technologies diversify, researchers have begun exploring how these constructs translate to other configurations. For example, work in augmented reality has examined how presence manifests when virtual and physical elements coexist, proposing early measurement approaches and adaptations of presence theory to AR contexts~\cite{regenbrecht2002measuring}.

\subsection{Measuring Presence and Immersion}
Alongside conceptual debates, a large body of work has investigated how immersive experience should be measured. Surveys of the literature show that questionnaires have dominated as the primary method, often administered post-experience~\cite{souza_2021}. Meta-analyses and reviews confirm that subjective scales are prevalent, while objective measures such as physiological signals remain underutilized~\cite{cummings_2016,bowman_2007}. Other studies point to the limitations of self-reports, advocating for neurophysiological measures~\cite{dey_2020} or continuous behavioral logging~\cite{robert_2024} as complementary approaches. Comprehensive reviews highlight the trade-offs of each method (e.g., questionnaires, task performance, physiological metrics, or hybrid approaches) and call for new ways of combining them~\cite{zhang_2020}. Despite decades of methodological innovation, we regret that no consensus has yet emerged on which measures best capture the elusive nature of immersion.

\subsection{Systematic Reviews of XR Evaluation Practices}
Several systematic reviews and surveys have mapped evaluation practices within specific XR domains. For AR research, early reviews categorized user evaluation techniques~\cite{dnser_2008} and analyzed usability studies across a decade of research~\cite{dey_2018}. Broader reviews of ISMAR proceedings identified trends in tracking, display, and interaction, noting a gradual shift toward user evaluation and realism~\cite{zhou_2008,kim_2018}. In VR, reviews have focused on user experience factors~\cite{kim_2020} and the impact of realism on outcomes~\cite{Goncalves_2022}. At the highest level, tertiary reviews synthesize findings from dozens of surveys, revealing that XR evaluation remains largely focused on effectiveness rather than efficiency~\cite{becker_2023}. Together, these reviews provide valuable insights, but they are not comprehensive, as they are either limited to a single technology or a single construct.

\subsection{Emerging Terminology}
More recent work has expanded the focus beyond presence to capture other facets of immersive experience. The sense of embodiment has been defined as a multidimensional construct comprising self-location, agency, and body ownership~\cite{Kilteni_2012}, with subsequent studies investigating how personal characteristics influence embodiment and task performance~\cite{peck_2024,li2024investigating,wang2025handows}. Similarly, systematic reviews have examined how VR affects empathy, finding consistent effects on cognitive but not emotional empathy~\cite{lee_2024}. Cognitive workload has also been reviewed, revealing heavy reliance on legacy measures like NASA-TLX and calling for better conceptualization within HCI~\cite{kosch_2023}. Other reviews highlight domain-specific applications, such as the effects of virtual nature on well-being~\cite{spano_2023} or the influence of frame rate on simulator sickness and performance~\cite{wang_2023}. Collectively, these works illustrate the growing recognition that immersive experience encompasses a broader set of constructs beyond presence alone.

\subsection{Frameworks and Methodologies for Evaluation}
In parallel, methodological frameworks have been proposed to systematize how immersive experiences are evaluated. Early work introduced taxonomies for analyzing navigation techniques~\cite{bowman_1998} or separating the effects of immersion from interaction techniques~\cite{McMahan_2006}. Other task-based approaches align user behaviors with multivariate performance metrics to provide fine-grained analyses~\cite{robert_2024}. Broader treatments of UX metrics advocate for their systematic adoption to guide design and improve reproducibility~\cite{albert2022measuring}. Across these contributions, the emphasis is on providing researchers with structured tools, yet these frameworks often address narrow dimensions of immersive interaction or focus on specific contexts~\cite{hale2014handbook,jerald2017human,laviola20173d}.

\section{Method}

\subsection{Search Strategy}

We conducted a literature search using the ACM Digital Library\footnote{\url{https://dl.acm.org/}} and IEEE Xplore\footnote{\url{https://ieeexplore.ieee.org/}}, focusing on high-impact and technical venues in HCI and related fields. Our strategy follows a bottom-up approach, analyzing how researchers describe user experiences with the interactive systems or experiments they propose. To achieve this, we selected the most reputable, influential, and representative venues within the HCI and XR communities. While this choice inevitably excludes some relevant works (e.g., PRESENCE: Virtual and Augmented Reality\footnote{\url{https://direct.mit.edu/pvar}} and Springer Virtual Reality\footnote{\url{https://link.springer.com/journal/10055}}), we expect that our review already covers a substantial number of empirical studies. Given our methodology, this selection is sufficient to summarize analytical trends in immersive virtual environments research.

\subsubsection{Keyword Identification}

We constructed a query to retrieve papers reporting immersive user experiences, adapting it to each database's search syntax. In the ACM Digital Library, for example, the query was:

\begin{quote}
\centering
(\textbf{TITLE} OR \textbf{ABSTRACT} (``immersi*'')) \\AND \textbf{ABSTRACT} (``experience'')
\end{quote}

This formulation captures variations such as ``immersion'' and ``immersive,'' while ensuring that ``experience'' appears in the abstract, indicating substantive engagement with user perception or evaluation. Due to differences in search functionalities between ACM DL and IEEE Xplore, the query syntax differs slightly, but we implemented the appropriate search expressions in each database to ensure an equivalent and consistent query across both.

\subsubsection{Database Selection and Query Execution}

Searches were conducted in the ACM Digital Library and IEEE Xplore, filtered to include only full-length research articles published before \textbf{December 1, 2024}. The venues targeted were:

\begin{itemize}
    \item \textbf{ACM CHI} (ACM CHI Conference on Human Factors in Computing Systems)
    \item \textbf{ACM UIST} (ACM Symposium on User Interface Software and Technology)
    \item \textbf{ACM VRST} (ACM Symposium on Virtual Reality Software and Technology)
    \item \textbf{ACM SUI} (ACM Symposium on Spatial User Interaction)
    \item \textbf{IEEE VR} (IEEE Conference on Virtual Reality and 3D User Interfaces)
    \item \textbf{IEEE ISMAR} (IEEE International Symposium on Mixed and Augmented Reality)
    \item \textbf{IEEE TVCG} (IEEE Transactions on Visualization and Computer Graphics)
\end{itemize}

\subsection{Preliminary Literature Screening}

To ensure both the relevance and the rigor of our results, we applied venue-specific preliminary filtering procedures prior to the full screening stage.

\textbf{ACM CHI and UIST.} Both CHI and UIST maintain main conference proceedings alongside companion extended abstract tracks. Using the ACM Digital Library’s advanced search tools, we filtered for records labeled as ``research articles'' and restricted the publication date to works published on or before December 1, 2024. Misclassified entries such as extended abstracts or posters erroneously tagged as full papers were manually identified and removed.

\textbf{ACM VRST and SUI}. Both VRST and SUI merged their short paper tracks (e.g., Demo, Posters, etc.) into their conference proceedings. Therefore, we also implemented the label ``research articles'' to distinguish the nature of our records. We also mutually identified and removed misclassified cases.

\textbf{IEEE VR and ISMAR.} Similarly, IEEE VR and ISMAR maintain separate proceedings for full papers and adjunct content. To ensure focus on complete research contributions, we restricted our selection to the full paper proceedings. However, in certain years (e.g., 2013, 2014), IEEE VR and ISMAR did not explicitly distinguish between adjunct and main paper proceedings. In such cases, we manually inspected and excluded short papers where applicable. Additionally, for IEEE VR 2012, all full papers were published in \textit{IEEE TVCG}; these were retrieved from the TVCG database and recorded under the IEEE VR label for consistency.

\textbf{IEEE TVCG.} As IEEE TVCG exclusively publishes peer-reviewed journal articles, all papers retrieved from this venue were assumed to meet our baseline research paper criterion.

\subsection{Review Process}

Our review process followed PRISMA-ScR guidelines~\cite{tricco2018prisma}. After removing duplicates, we identified 721 articles. We conducted two rounds of screening: (1) title, abstract, and keyword screening, and (2) full-text screening. The first and second authors independently reviewed all records in both rounds. Disagreements were resolved through discussion, and when necessary, by consulting a senior expert. Prior to the full review, the coding scheme and decision criteria were jointly developed and piloted on a subset of papers to ensure a shared understanding. Consistency checking was carried out by comparing independently coded samples at multiple points during the process, which helped maintain alignment and reduce subjective drift. We applied the following exclusion criteria (EC) to eliminate irrelevant papers and included papers if they met at least one of the inclusion criteria (IC):

\textbf{EC1.} \textit{Not in main proceedings}: The papers not in main proceedings were excluded, such as keynote papers, posters, extended abstracts, short papers (usually within 3 pages), and workshop proposals---as these formats generally do not provide sufficient space to clearly describe the system or the evaluation methods used, and many represent work-in-progress rather than fully developed studies.

\textbf{EC2.} \textit{False positive}: Exclude papers where the claimed immersive context is misleading or inconsistent with the actual focus. For instance, if a work introduces a new algorithm or back-end system but only loosely frames it as ``immersive'' without providing substantial user-facing experience or interaction design, it is considered a false positive and removed.

\textbf{EC3.} \textit{Example mention}: Exclude papers in which immersive technology is only briefly or superficially mentioned (e.g., a passing reference in the abstract or introduction) and does not form the central subject of study. This ensures that only papers offering substantive systems, applications, or analyses of immersive experience are retained.

\textbf{EC4.} \textit{Literature review or survey}: Surveys and literature reviews are excluded. The purpose of this study is to analyze the specific techniques, systems, measurements, and original research, not to review existing review studies. 

\textbf{IC1.} \textit{Application focus}: These are studies where the application of technology is not the central research objective but plays a crucial role in other fields such as training, rehabilitation, and learning. An example would be a project leveraging augmented reality (AR) to enhance education and interaction.

\textbf{IC2.} \textit{Innovative interaction techniques}: Papers should present a novel interaction technique, and evaluate the technique by conducting user studies, especially highlighting the comparison with other typical interaction techniques. This includes but is not limited to, selection, manipulation, teleportation and navigation.

\textbf{IC3.} \textit{User experience and system usability}: It includes studies that analyze and compare implicit, subjective, or potential factors contributing to variations in user experience. For instance, the paper would include exploring what influences motion sickness and other underlying elements that shape user perception and comfort.

\textbf{IC4.} \textit{Comprehensive evaluation of systems}: Consider those studies that propose or conduct holistic assessments of systems. This may involve evaluating the system as a whole rather than focusing on individual interaction technologies, emphasizing overall coherence and integration.

\textbf{IC5.} \textit{Articles requiring further reading}: If, at the time of initial reading of the abstract, it is not clear whether the articles meet the inclusion criteria, we included them for the second stage of screening.

\subsubsection{Screening Rounds}

\paragraph{\textit{First round.}} In this stage, we first excluded 17 papers by screening papers based on page count, and categories identified (EC4). Although we used the integrated filters to screen some records from databases, there are still some papers left due to the problems with these filters. Then, 298 records were screened by the authors (EC1, EC2, and EC3) based on title and abstract, resulting in a pool of 406 papers included for the second round of screening.

\begin{figure}
    \centering
    \includegraphics[width=\linewidth]{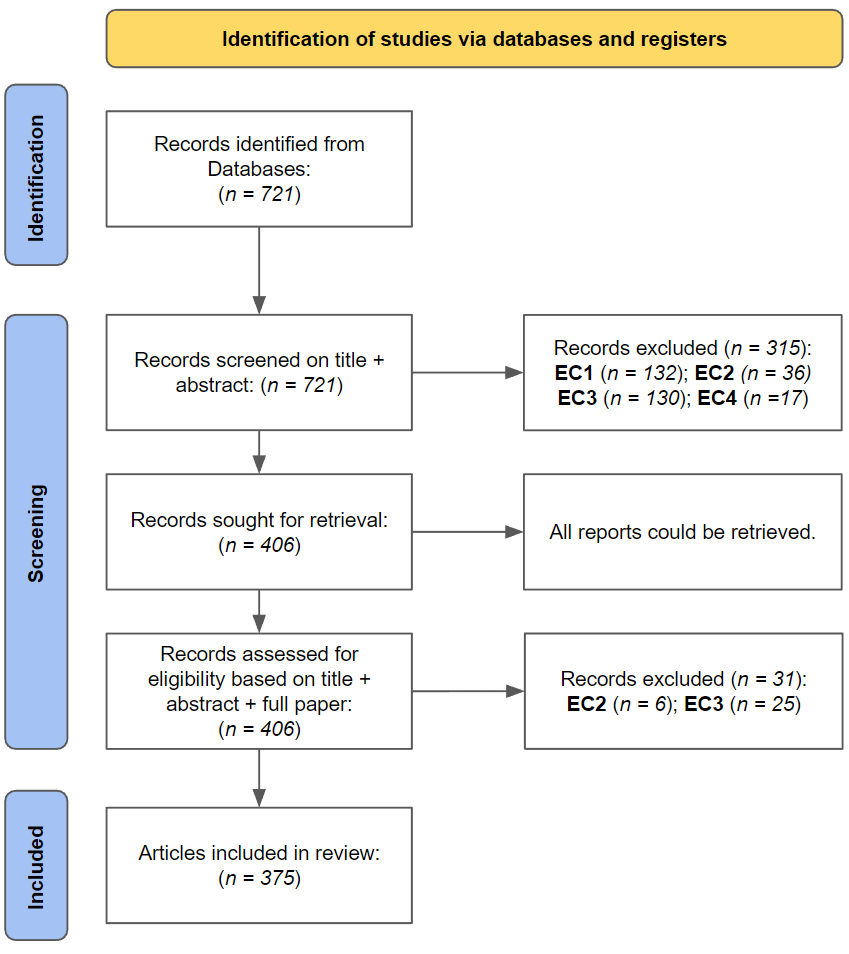}
    \caption{PRISMA flowchart detailing the review process for identifying, screening, and including studies in the review. A total of 721 records were identified from database searches. After screening titles and abstracts, 315 records were excluded based on exclusion criteria (EC1–EC4). Of the 406 records sought for full-text retrieval, all were retrieved successfully, with 31 further excluded following full-paper assessment. Ultimately, 375 articles were included in the final review.}
    \Description{PRISMA flowchart detailing the review process for identifying, screening, and including studies in the review. A total of 721 records were identified from database searches. After screening titles and abstracts, 315 records were excluded based on exclusion criteria (EC1–EC4). Of the 406 records sought for full-text retrieval, all were retrieved successfully, with 31 further excluded following full-paper assessment. Ultimately, 375 articles were included in the final review.}
    \label{fig:prisma}
\end{figure}

\paragraph{\textit{Second round.}} We conducted full-text reviews of the remaining 406 papers. Applying the same EC/IC criteria, we excluded an additional 31 papers, resulting in a final dataset of 375 full research papers. The complete PRISMA flow diagram is shown in \autoref{fig:prisma}.

\section{Results}

\begin{figure*}
    \centering
    \includegraphics[width=\linewidth]{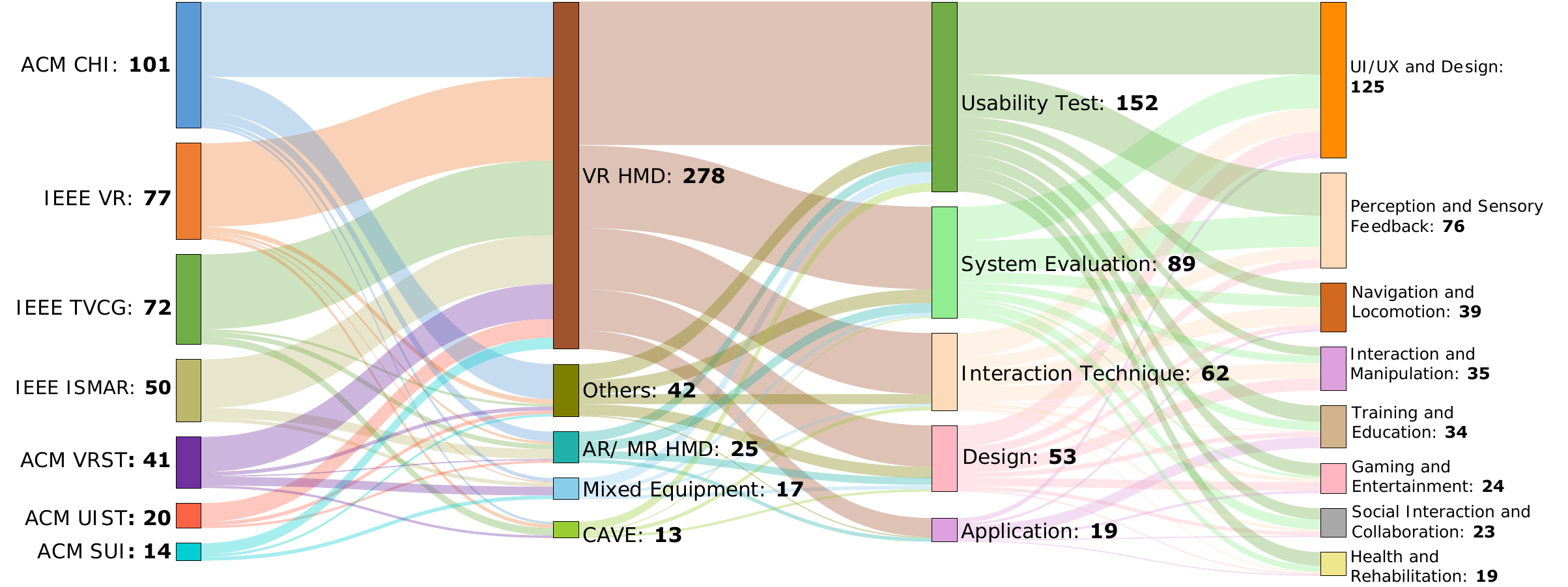}
    \caption{Overview of the reviewed literature on immersive interaction technologies, categorized by publication venues, device types, primary contributions, and application domains. The diagram illustrates the flow of research contributions from key venues (IEEE TVCG, IEEE VR, IEEE ISMAR, ACM CHI, ACM UIST, ACM VRST and ACM SUI) through device categories (e.g., VR, AR/MR, CAVE systems) and primary contributions (e.g., usability testing, system evaluation, interaction techniques) to specific application domains (e.g., UI/UX, navigation, training, social interaction).}
    \Description{A Sankey diagram visualizing the flow of research contributions in immersive interaction technologies, categorized by publication venues, device types, primary contributions, and application areas. The diagram shows research contributions from major venues, including IEEE TVCG, IEEE VR, IEEE ISMAR, ACM CHI, ACM UIST, ACM VRST, and ACM SUI, which are linked to device categories such as VR, AR/MR, CAVE systems, and mixed equipment. These devices are further mapped to primary contributions like usability testing, system evaluation, and interaction techniques. Finally, the research is connected to specific application domains, including UI/UX and design, gaming and entertainment, health and rehabilitation, navigation and locomotion, social interaction and collaboration, training and education, interaction and manipulation, and perception and sensory feedback.}
    \label{fig:sankey}
\end{figure*}

This section presents the results of our scoping review of 375 full research papers published over the last 30 years (i.e., between 1995 and 2024), across seven premier venues in HCI and immersive technologies: \textit{ACM CHI} (101 papers), \textit{IEEE VR} (77), \textit{IEEE TVCG} (72), \textit{IEEE ISMAR} (50), \textit{ACM VRST} (41), \textit{ACM UIST} (20) and \textit{ACM SUI} (14) (see \autoref{fig:four_pies} (A)). We first provide a high-level overview of dataset characteristics (also see \autoref{fig:sankey}). We then describe our coding strategy and classification principles, including deviations and edge cases encountered. Finally, we present detailed findings across the following analytical dimensions: venue distribution and temporal trends, device types, research contribution types, application domains, evaluation method combinations, and questionnaire usage and constructs.

\subsection{Overview of Dataset Characteristics}

Our final dataset spans 30 years since 1995, reflecting both the technological progression of immersive systems and an increasing scholarly focus on evaluating experience-related phenomena. \autoref{fig:line} illustrates the annual publication trends. The number of relevant papers remained low and relatively stable until the mid-2010s. Beginning in 2013, we observed a sharp increase in publication volume, with an average of over 30 papers published annually after 2018. This growth reflects the combined impact of hardware accessibility, the maturity of XR development platforms, and the rising importance of user-centered design and evaluation in immersive system research.

\subsection{Coding Procedure and Classification Principles}

To guide our analysis, we established a multi-dimensional coding scheme comprising \texttt{device modality}, primary \texttt{contribution type}, and \texttt{application domain}. In addition, we specified categories for \texttt{evaluation methods}. All coding was conducted manually by the authors through iterative review, with discrepancies resolved by discussion and consensus.

\begin{acmbox}{i. Device Modalities}
We categorized papers according to the immersive context in which the interaction technique or system was deployed. 
\end{acmbox}

\begin{figure*}[t]
    \centering
    \includegraphics[width=0.9\linewidth]{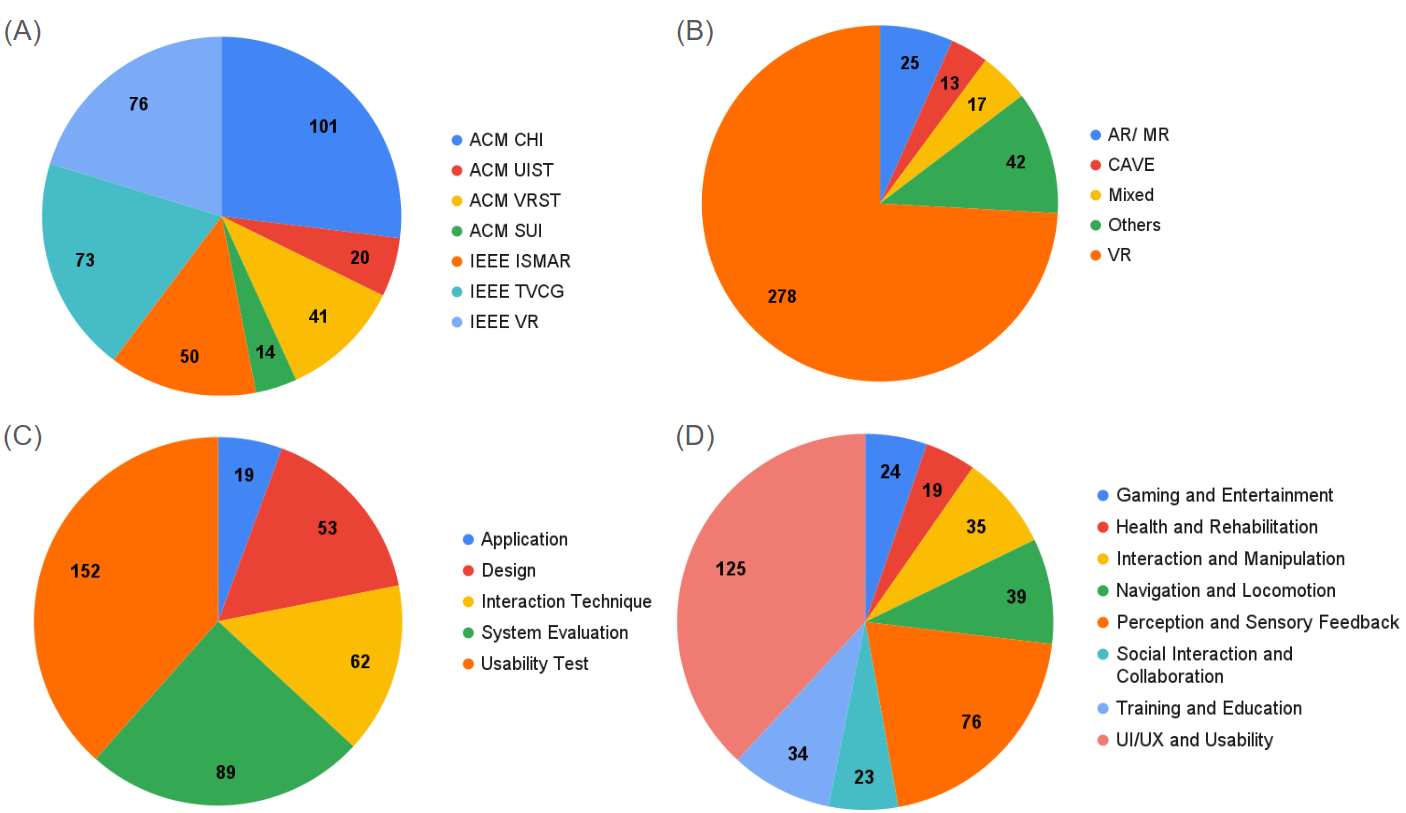}
    \caption{Categorization of studies based on four classification methods. The total number of studies is consistent across all charts, which categorize the data according to the following criteria: (a) Conference, (b) Equipment, (c) Primary Contribution, and (d) Application Domain.}

    \Description{Four pie charts categorizing the 375 reviewed studies across different criteria. 
    Chart (a) shows the distribution by Conference: ACM CHI (101 studies), ACM UIST (20), ACM SUI (14), ACM VRST (41), IEEE ISMAR (50), IEEE TVCG (72), and IEEE VR (77). 
    Chart (b) categorizes studies by Equipment: AR/MR (25), CAVE (13), Mixed (17), Others (42), and VR (278). 
    Chart (c) classifies them by Primary Contribution: Application (19), Design (53), Interaction Technique (62), System Evaluation (89), and Usability Test (152). 
    Chart (d) groups them by Application Domain: Gaming and Entertainment (24), Health and Rehabilitation (19), Interaction and Manipulation (35), Navigation and Locomotion (39), Perception and Sensory Feedback (76), Social Interaction and Collaboration (23), Training and Education (34), and UI/UX and Design (125).}
    \label{fig:four_pies}
\end{figure*}

For example, if a study examined a haptic system used within a virtual environment, it was labeled as \texttt{VR}. When the same technique was implemented and evaluated in multiple distinct conditions, it was labeled as \texttt{Mixed} (see Figure~\ref{fig:four_pies} (B) for the distribution). This scenario-driven categorization provides a more ecologically valid understanding of how systems are situated in practice.

\begin{acmbox}{ii. Primary Contributions}
We assigned each paper a single label reflecting its primary contribution type. The categories include: \texttt{Usability Test}, \texttt{System Evaluation}, \texttt{Interaction Technique}, \texttt{Design}, and \texttt{Application}. Each label was determined based on the paper's central research aim and methodological structure. 
\end{acmbox}

While many papers span multiple contribution types, we consistently selected the most relevant type through author consensus. For example, a study that develops a domain-specific immersive system (e.g., an educational MR app) with its main emphasis on use case implementation, rather than introducing new interaction techniques, was labeled as \texttt{Application}. \texttt{Interaction Technique} refers to work proposing and evaluating a novel method of interaction within VR/AR/MR (e.g., pointing, text entry, locomotion), with contributions lying in the technique itself and its evaluation. \texttt{System Evaluation} refers to research that assesses an entire system or platform, in terms of performance, feasibility, or comparative benchmarking, often at an infrastructural or technical level (see Figure~\ref{fig:four_pies} (C)).


\begin{acmbox}{iii. Application Domains}
We categorized each paper into one specific domain depending on its empirical context or target scenario. The domains include: \texttt{UI/UX and Design}, \texttt{Navigation and Locomotion}, \texttt{Training and Education}, \texttt{Perception and Sensory Feedback}, \texttt{Interaction and Manipulation}, \texttt{Health and Rehabilitation}, \texttt{Social Interaction and Collaboration}, and \texttt{Gaming and Entertainment}.
\end{acmbox}

This domain-based classification approach highlights the breadth of application areas in which immersive systems are deployed (see Figure~\ref{fig:four_pies} (D)). It helps to situate contributions within concrete use cases, making it easier to compare work across distinct contexts (e.g., contrasting systems designed for education versus those targeting health or entertainment).

\begin{acmbox}{iv. Evaluation Methods}
We identified five non-exclusive categories: \texttt{standardized questionnaires}, \texttt{task (or participants' behavioral) performance}, \texttt{system performance metrics}, \texttt{interviews (or open-ended responses)}, and \texttt{physiological measures}. Papers were also annotated for the number of method types used.
\end{acmbox}

This classification highlights the methodological diversity in evaluating immersive systems. By capturing both subjective (e.g., questionnaires, interviews) and objective (e.g., performance, physiological, system metrics) approaches, it provides insight into how rigor and triangulation are achieved across studies.

\subsection{Venue Distribution and Temporal trends}
The 375 papers in our review are distributed across \textit{ACM CHI} (101 papers), \textit{IEEE VR} (77), \textit{IEEE TVCG} (72), \textit{IEEE ISMAR} (50), \textit{ACM VRST} (41), \textit{ACM UIST} (20) and \textit{ACM SUI} (14). Together, these venues mainly represent both conference- and journal-style publication models and capture the interdisciplinary breadth of HCI and XR research.

\begin{figure}
    \centering
    \includegraphics[width=\linewidth]{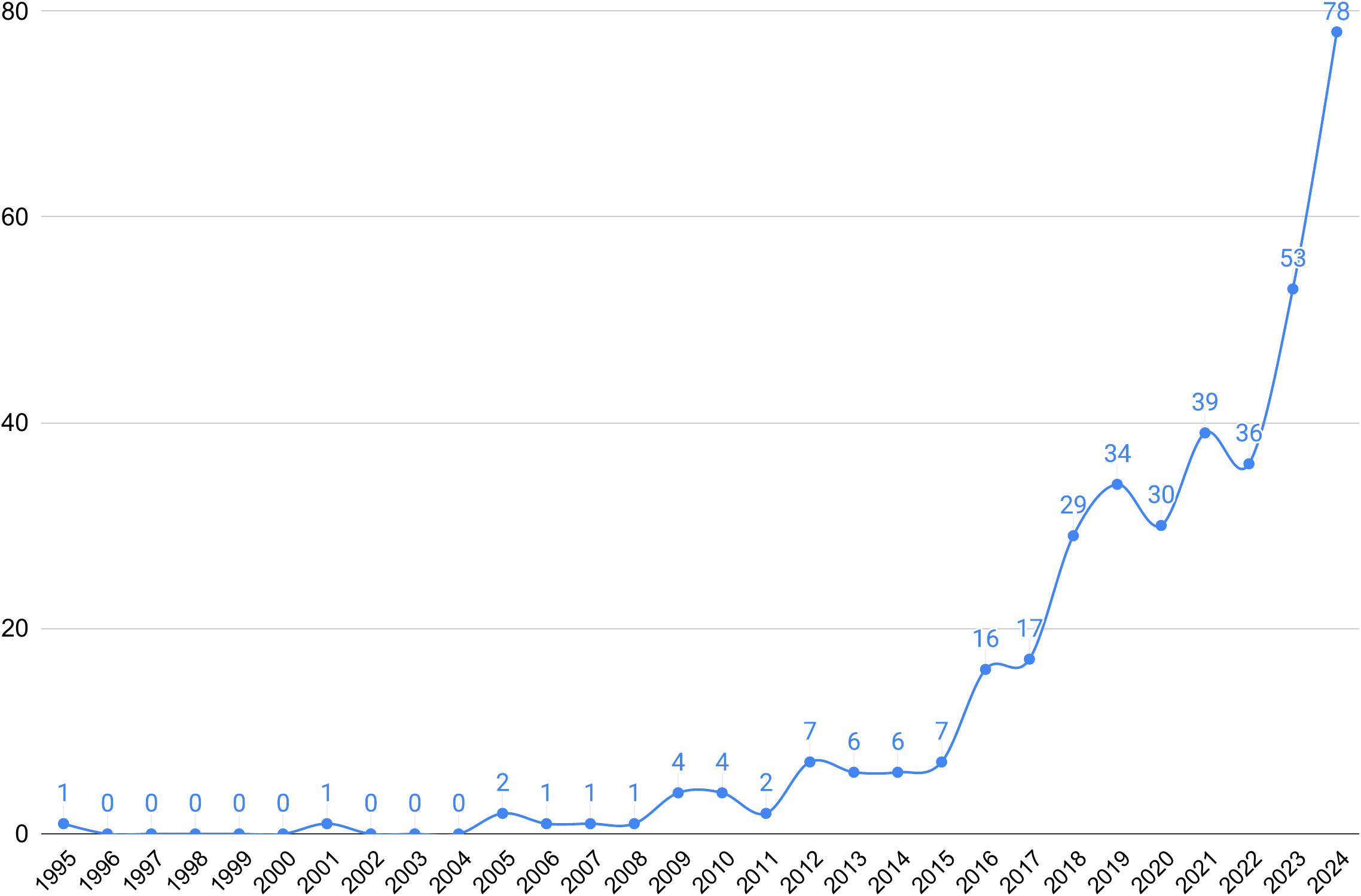}
    \caption{Annual distribution of publications from 1995 to 2024. Our results illustrate a significant upward trend in research interest over the last decade, with an exponential surge in publications observed in 2023 and 2024.}
    \Description{
    A line chart showing the number of publications per year from 1995 to 2024. The chart starts with very low numbers from 1995 to around 2008, mostly between 0 and 2. From 2009 to 2015, the values fluctuated slightly between 1 and 7 publications. A noticeable increase starts in 2016 with 9 publications, rising sharply to 17 in 2017, 29 in 2018, and peaking at 34 in 2019. The numbers remain relatively stable between 2020 and 2022, ranging from 30 to 39. In 2024, there is a sharp rise to 78 publications, the highest in the entire range.}
    \label{fig:line}
\end{figure}

\autoref{fig:line} illustrates the annual distribution of publications in our dataset from 1995 to 2024. These numbers reflect only the papers included in our review and should not be interpreted as the total volume of immersive experience research produced during this period; rather, they indicate the relative trends captured by our sampling strategy. For the first 15 years, immersive experience research appeared only intermittently in our dataset, with publication counts rarely exceeding one or two papers per year. Beginning around 2011, however, we observed a steady rise, followed by a pronounced acceleration after 2017. The number of papers increased from 16–17 in 2016–2017 to nearly 30 by 2018, continuing upward through the early 2020s and culminating in 53 papers in 2023 and 78 in 2024. This trend aligns with broader developments in the XR ecosystem: the maturation of consumer VR hardware, the widespread adoption of accessible development platforms, and growing methodological emphasis within HCI and XR communities on experience-centered evaluation. Together, these factors contribute to the strong upward trajectory observed in our dataset.

Across venues, we observe that CHI consistently accounts for the largest number of papers. IEEE VR and ISMAR, as premier XR-focused venues, exhibit steady contributions with some recent growth. TVCG, although a journal venue, contributes a substantial volume of immersive UX research, particularly via special issues and conference-journal integration models (e.g., for IEEE VR 2012, all accepted papers were published at IEEE TVCG). UIST contributes a smaller but consistently high-quality set of papers, often focused on technical innovation in interaction techniques. Together, these trends suggest that immersive experience research has matured into a robust area of inquiry, with sustained output and methodological diversification across multiple publication formats.

\section{Approaches to Measure the Immersive Experience}

\begin{figure}[t]
    \centering
    \includegraphics[width=\linewidth]{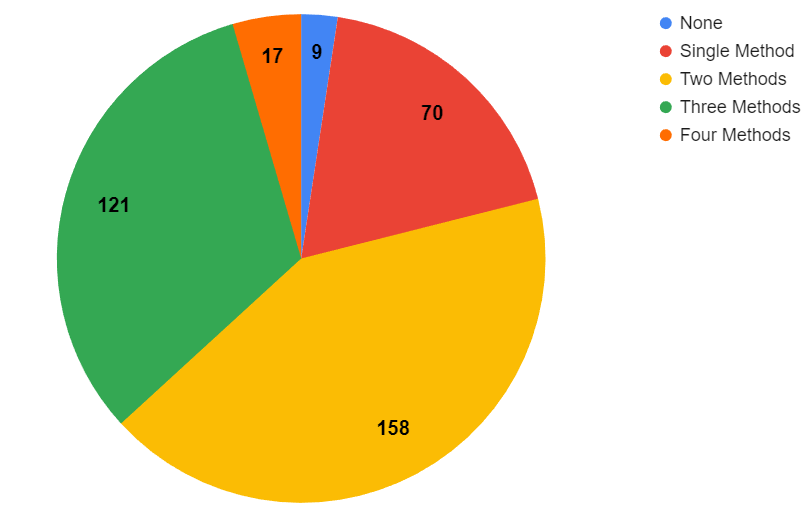}
    \caption{Proportion of papers by number of UX evaluation methods used. The majority of studies adopted either two methods (\(N = 158\), 42.1\%) or three methods (\(N = 121\), 32.3\%), suggesting a growing preference for mixed-method evaluation strategies. A smaller portion of papers relied on only one method (\(N = 70\), 18.7\%), while only 4.5\% used four methods (\(N = 17\)). Notably, 2.4\% of papers (\(N = 9\)) did not report any user experience evaluation. No paper in our dataset used all evaluation methods.}
    \Description{Pie chart showing proportions of immersive systems papers by number of UX evaluation methods used: 42.1\% (158 papers) used two methods, 32.3\% (121) used three, 18.7\% (70) used one, 4.5\% (17) used four, and 2.4\% (9) used none. No paper in our dataset used all evaluation methods.}
    \label{fig:evaluation_method_set_size}
\end{figure}

Our synthesis identified five primary approaches used to evaluate immersive experience: \texttt{Questionnaires}, \texttt{System Performance}, \texttt{Task Performance}, \texttt{Interviews}, and \texttt{Physiological Measures}. Each approach captures distinct facets of user experience. Questionnaires ($N=321$, 85.6\%) were by far the most common, offering structured access to psychological constructs such as presence, engagement, and workload. Task performance measures ($N=235$, 62.7\%), such as completion time or error rates, and system performance measures ($N=52$, 13.9\%), such as latency or frame rate, instead quantify immersion indirectly through observable outcomes and technical indicators. Interviews ($N=169$, 45.1\%) provided qualitative depth, foregrounding user perspectives and contextual nuances often absent from structured scales. Physiological measures ($N=40$, 10.7\%) (e.g., heart rate variability, electrodermal activity, and eye tracking), enabled researchers to probe embodied and affective responses not easily verbalized. Only a very small number of papers ($N=9$, 2.4\%) reported no explicit user experience metric at all (see \autoref{fig:evaluation_method_set_size}.) 

Importantly, this methodological expansion reflects the changing nature of XR research itself. Earlier work, particularly in 2016 and before, often focused on establishing core system implementations, which meant there were fewer application templates and consequently fewer opportunities or expectations for rich experience-oriented evaluation. In contrast, contemporary immersive applications are more diverse, detailed, and mature, which has enabled and motivated the adoption of a broader range of approaches for capturing immersive experiences.

\subsection{Relying on Questionnaires}

We documented questionnaire use with particular attention to source validity, distinguishing between standardized instruments (i.e., those with clearly defined constructs and published references) and self-developed questionnaires without formal validation. In cases where citation practices were ambiguous or incomplete, we applied a conservative classification strategy. For instance, ~\cite{kruzan_supporting_2020} referenced the use of \texttt{GAD7} and \texttt{PHQ9} to assess anxiety and depression symptoms, yet did not provide appropriate citations or explanatory details. We treated such instances as citation omissions and classified the instruments as self-developed. Similarly, we identified citation inaccuracies, such as~\cite{kim2024audience} referenced a review article on questionnaires~\cite{norman2013geq}, rather than the validated GEQ instrument~\cite{BROCKMYER2009624}. 

In cases involving multiple versions of a questionnaire, we consolidated references where appropriate. For the State–Trait Anxiety Inventory (STAI), we identified five papers in our dataset that used it to measure anxiety. Four of them \cite{kern_immersive_2019,min_effects_2020,dickinson_experiencing_2021,kwon2022infinite} cited the original STAI source \cite{Spielberger1970STAI} (Form X), while one paper \cite{Zhang2023Forestlight} cited a meta-analysis \cite{KNOWLES2020101928}. Upon examining this meta-analysis, we found that Spielberger et al. later released a revised version of the STAI \cite{Spielberger_1983_STAI}, known as Form Y. The revision was motivated by the concern that several items in Form X overlapped with depressive symptoms (e.g., ``I feel blue''), and thus Form Y was developed to improve content validity. Meta-analytic findings further indicate that Form Y produces significantly larger effect sizes than Form X when differentiating individuals with anxiety disorders from healthy controls, suggesting higher specificity \cite{KNOWLES2020101928}. Another example here is that, ~\cite{xu_2019_pointing} cited the original \texttt{UEQ}~\cite{laugwitz_construction_2008}, while~\cite{giovannelli2022exploring} employed the \texttt{UEQ-Short}~\cite{schrepp2017design}. As the short version is an explicitly derived update that cites the original scale, we merged their instances and retained both references in~\autoref{tab:questionnaire_summary}, which lists the most frequently used questionnaires that were referenced at least \textbf{three} times in our dataset.

\begin{table*}[t!]
\caption{Summary of Questionnaires, Paper Counts, and Corresponding Papers for Various Dimensions of User Experience in Virtual Environments.}
\label{tab:questionnaire_summary}

\footnotesize 

\begin{tabularx}{\textwidth}{p{2.5cm} l c X}
\toprule
\textbf{Main Dimension} & \textbf{Questionnaire} & \textbf{Count} & \textbf{Paper} \\
\midrule

\textbf{Sickness} & & \textit{\textbf{88}} & \\
& SSQ \cite{kennedy_simulator_1993} & 82 &  \cite{young2007demand,young2006demand,suma2011leveraging,schild_understanding_2012,bolte2015subliminal,freitag_examining_2016,schatzschneider_who_2016,mcgill_i_2017,hock_carvr_2017,lin_tell_2017,kasahara2016jackin,naz_emotional_2017,knibbe_dream_2018,lee2018user,schmitz2018you,lugrin2018any,grubert2018text,sra_adding_2019,latoschik_not_2019,kang_jumping_2019,lee_simulating_2019,kern_immersive_2019,hamzeheinejad_physiological_2019,wolf_jumpvr_2020,yeo_toward_2020,monteiro2020depth,nie2019analysis,venkatakrishnan2020comparative,Venkatakrishnan_2020_Equation,liao2020data,bala_dynamic_2021,feick_visuo-haptic_2021,Li_2021_evaluating,nakano_head-mounted_2021,jung_floor-vibration_2021,zenner2021combining,cmentowski_towards_2021,singla_assessment_2021,vaziri_egocentric_2021,kwon2022infinite,pointecker2022bridging,groth_omnidirectional_2022,hashemian_headjoystick_2022,kim_studying_2022,liu2022investigating,amaro2022design,martin2023study,hashemian_leaning-based_2023,luan_2024_exploring,lu_2024_veer,mahmud2024multimodal,xu2024saferdw,xu2024spatial,joo2024effects,tian2024enhancing,liu2024visioncoach,Friedl2024collaboration,merz2024universal,zou2024effect,Venkatakrishnan_2024_effects,guo2024breaking,pohlmann_you_2023,mimnaugh2023virtual,mahmud2023auditory,hwang_electrical_2023,hwang_2023_enhancing,zhao2023leanon,Zielinski2013Intercept,Chen20136dof,Llorach2014Simulator,Ariza2016Inducing,Arafat2016Effects,Westhoven2016Head,Argelaguet2016Giant,Valkov2017Smooth,Chowdhury2017Information,Min2020Effects,Lee2021Design,Podkosova2023Joint,Xu2024Istraypaws,Holly2024Gamebased,Kim2024Lifter} \\
& VRSQ \cite{kim_virtual_2018} & 6 & \cite{singla_assessment_2021, bala_dynamic_2021, zhu2023can, xu2023user,monty2024analysis,Landeck2023From} \\
\midrule
\textbf{Engagement} & & \textit{\textbf{9}} & \\
 & NES \cite{busselle_measuring_2009} & 5 &  \cite{mert_canat_sensation_2016, bindman_am_2018, shin_user-oriented_2021,shin2023effects,belz2024story} \\
& GEQ \cite{BROCKMYER2009624} & 4 &  \cite{mcmahan_evaluating_2012,rogers_exploring_2019,kim2024audience,rogers_vanishing_2018} \\
\midrule
\textbf{Task Load} & & \textit{\textbf{62}} & \\
& NASA-TLX \cite{hart_development_1988, hart2006} & 56 &  \cite{gandy2010experiences,chalil2011synchronous,bertrand_role_2015,freitag_examining_2016,Westhoven2016Head,boldt2018you,grubert2018text,xu_2019_pointing,hamzeheinejad_physiological_2019,choi_virtual_2019,li2020analysing,cai2020thermairglove,gerling_virtual_2020,gerling_virtual_2020,cmentowski_towards_2021,hsu_comparative_2021,dickinson_diegetic_2021-1,shin2021user,Li_2021_evaluating,Lee2021Design,Safikhani2021Influence,hashemian_headjoystick_2022,mathis2022virtual,giovannelli2022exploring,thoravi2022dreamstream,amaro2022design,liu2022investigating,Shi2022Groupbased,belani2023investigating,venkatakrishnan2023virtual,shin2023effects,xu2023gesturesurface,pohlmann_you_2023,Podkosova2023Joint,Ullah2023Exploring,zou2024effect,Lee2024Viewer2Explorer,Wang2024exploring,seo2024gradualreality,He2024learning,wang2024a3rt,Friedl2024collaboration,Luo2024Exploring,Kim2024Crossing,Song2024Exploring,Simon2024impact,Acevedo2024effects,Ilo2024Goldilocks,Eroglu2024Choose,Kim2024Lifter,cauquis2024investigating,monty2024analysis,kruger2024intenselect,merz2024voice,merz2024universal,Gabel2023Redirecting} \\
& VHP \cite{mal2022virtual} & 3 &  \cite{merz2024voice,merz2024universal,Arboleda2024beyond} \\
& SMEQ \cite{zijlstra1985construction} & 3 &  \cite{lee2018user, thanyadit2019observar, gattullo2022biophilic}  \\
\midrule
\textbf{Presence} & & \textit{\textbf{138}} & \\
 & IPQ \cite{schubert_experience_2001} & 41 &  \cite{mcgill_i_2017,wilberz_facehaptics_2020,wolf_jumpvr_2020,jain_immersive_2016,buttussi_effects_2018,kanamori2018obstacle,grubert2018text,kang_jumping_2019,gunther_effect_2019,valentini2020improving,peng_palette_2020,bala_dynamic_2021,
 dickinson_diegetic_2021-1,nakano_head-mounted_2021,malinov_stuckinspace_2021,endo_modularhmd_2021,he_context-consistent_2022,tatzgern_airres_2022,mathis2022virtual,pohlmann_you_2023,hwang_electrical_2023,kocur_effects_2023,kim2023or,han_architectural_2023,schott2023vreal,seo2024gradualreality,joo2024effects,Villa2024touch,chen2024video2haptics,merz2024voice,lee2024speak,liu2024visioncoach,merz2024universal,tseng2024understanding,zhang_2024_jump,Westhoven2016Head,Ratcliffe2022Rich,Somarathna2023Exploring,Landeck2023From,Xu2024Istraypaws,Kim2024Lifter} \\
& PQ \cite{witmer_measuring_1998} & 40 &  \cite{luan_2024_exploring,mcquiggan_effects_2008,gandy2010experiences,chalil2011synchronous,Chen20136dof,Llorach2014Simulator,bertrand_role_2015,jain_immersive_2016,mert_canat_sensation_2016,Bozgeyikli2016Locomotion,lee_exploring_2017,peiris_thermovr_2017,bindman_am_2018,ghosh_notifivr_2018,rietzler_breaking_2018,Skarbez2018Immersion,muender_does_2019,cheng_vroamer_2019,lee_simulating_2019,li2020analysing,cai2020thermairglove,ricca2020influence,dey2020neurophysiological,skarbez2020immersion,yeo_toward_2020,Min2020Effects,cmentowski_towards_2021,shin_user-oriented_2021,worrallo2021robust,Lee2021Design,tseng_headwind_2022,drey_towards_2022,tao_integrating_2022,chen2023leap,shin2023effects,kim2024QuadStretcher,Li_2021_evaluating,kim2024audience,Lee2024Viewer2Explorer,fan2024spinshot} \\
 & SUS (Presence) \cite{usoh_using_2000} & 31 &  \cite{heldal2005immersiveness,steinicke_does_2009,suma2011leveraging,mcmahan_evaluating_2012,freitag_examining_2016,schatzschneider_who_2016,Ariza2016Inducing,gugenheimer_sharevr_2017,sra_breathvr_2018,Skarbez2018Immersion,xu_2019_pointing,choi_virtual_2019,hartmann_realitycheck_2019,liao2020data,gerling_virtual_2020,dey_2020,skarbez2020immersion,venkatakrishnan2020comparative,hsu_comparative_2021,worrallo2021robust,zenner2021combining,Gauthier2021Virtual,hashemian_headjoystick_2022,woodward_it_2022,woodward_it_2022,kang2023giant,hashemian_leaning-based_2023,covaci_multisensory_2023,xiong2024petpresence,Bonnail2024real,Cools2024Impact} \\
 & NMMoSP \cite{Biocca2001TheNM,harms2004internal} & 10 &  \cite{malinov_stuckinspace_2021, yang_towards_2022, merz2024universal,jing2024superpowering,merz2024voice,Immohr2024evaluating,guo2024breaking,liu2022investigating,chen2024effects,thanyadit2019observar} \\
 & MEC-SPQ \cite{vorderer_mec_2004} & 6 &  \cite{lee2018user,zhao_scale_2019, rieger_how_2023,jung_floor-vibration_2021,zhao2023leanon,schild_understanding_2012} \\ 
   & Sense of presence, social presence and co-presence \cite{nowak_2003_presence} & 4 & \cite{wang_2024_real,Simon2024impact,parmar_how_2023,parmar_programming_2016} \\
 & SPQ \cite{jeremy_2003_spatial} & 3 & \cite{lee_exploring_2017,Garcia2024speaking,Arboleda2024beyond} \\
  & ITC-SOPI \cite{Lessiter2001ITC} & 3 &  \cite{dickinson_experiencing_2021,raeburn2021varying,bialkova_encouraging_2019} \\
  \midrule
\textbf{Emotion} & & \textit{\textbf{39}} & \\

& IMI \cite{ryan_self-determination_2000} & 11 &  \cite{wolf_jumpvr_2020, boldt2018you, harris_asymmetry_2019, kern_immersive_2019, bialkova_encouraging_2019,hamzeheinejad_physiological_2019,emmerich_streaming_2021,cmentowski_towards_2021,kao_audio_2022,salagean2024watch,Ratcliffe2022Rich} \\
& SAM \cite{bradley_measuring_1994} & 7 &  \cite{canat_sensation_2016, kern_immersive_2019, wilberz_facehaptics_2020, altmeyer_here_2022,jing2024superpowering,chen2024effects,tseng2024understanding} \\
& PENS \cite{ryan_motivational_2006} & 7 &  \cite{johnson_all_2015, harris_asymmetry_2019, gerling_virtual_2020, kao_audio_2022,guo2024breaking,Jung2023Crossreality,Holly2024Gamebased} \\
& PANAS \cite{watson1988development} & 6 &  \cite{kim2023active, dey_effects_2017, rogers_vanishing_2018, krum2018influences, boldt2018you, wagener_mood_2022} \\ 
& STAI \cite{Spielberger1970STAI} & 5 &  \cite{kern_immersive_2019,min_effects_2020,dickinson_experiencing_2021,kwon2022infinite} \\
 & I-PANAS-SF \cite{thompson_development_2007} & 3 &  \cite{hamzeheinejad_physiological_2019, kwon2022infinite,kern_immersive_2019} \\


\midrule
\textbf{Experience} & & \textit{\textbf{42}} & \\
& UEQ \cite{laugwitz_construction_2008} & 16 &  \cite{xu_2019_pointing,gattullo2022biophilic,schott2023vreal,xu2023user,hamzeheinejad_physiological_2019,gunther_effect_2019,zenner_immersive_2020,liu2022investigating,belani2023investigating,Garcia2024speaking,Kim2024Crossing,kruger2024intenselect,Friedl2024collaboration,kern_immersive_2019,Xu2024Istraypaws,Eroglu2024Choose} \\
& GEQ \cite{ijsselsteijn_game_2013,Poels2007geq_book} & 11 &  \cite{schild_understanding_2012,johnson_all_2015,mert_canat_sensation_2016,ju2016personality,gugenheimer_sharevr_2017,sra_breathvr_2018,sra_adding_2019,shin_user-oriented_2021,liu2022investigating,Jung2023Crossreality,Christiansen2024Exploring} \\ 
& E$^2$I  \cite{lin_effects_2002} & 6 &  \cite{hock_carvr_2017, rietzler_breaking_2018, rogers_exploring_2019, lee_simulating_2019,kang_jumping_2019,kwon2022infinite} \\
& UEQ-Short  \cite{schrepp2017design} & 5 &  \cite{giovannelli2022exploring,pointecker2022bridging,Song2024Exploring,belz2024story,zhang_2024_jump} \\
& PXI \cite{abeele2020development,vanden2016design} & 4 &  \cite{wolf_jumpvr_2020,drey_towards_2022,cmentowski_towards_2021,kao_audio_2022} \\ 
\midrule

\textbf{Immersion} & & \textit{\textbf{32}} & \\
& IEQ \cite{jennett_measuring_2008} & 14 &  \cite{cox_not_2012, cairns_influence_2014, canat_sensation_2016, lugrin2018any,rogers_vanishing_2018,rogers_exploring_2019,  peng_palette_2020,monteiro2020depth, kim_studying_2022, mella_gaming_2023, kim2023or, drey_towards_2022,He2024learning,He2024learning} \\
& AEQ \cite{gonzalez-franco_avatar_2018, peck2021avatar} & 11 &  \cite{schmitz2018you,fang_retargeted_2021,Buck2022TheIO,tatzgern_airres_2022,bhargava2023empirically,ganias2023comparing,kim2023or,venkatakrishnan2023virtual,cheng2024first,kim2024selfavatar,peck2024measuring} \\
& ITQ \cite{witmer_measuring_1998} & 7 &  \cite{mcquiggan_effects_2008,m_e_latoschik_not_2019,mert_canat_sensation_2016,schmitz2018you,liao2020data,raeburn2021varying,merz2024universal} \\

\midrule
\textbf{Usability} & & \textit{\textbf{26}} & \\
& SUS (Usability) \cite{brooke_sus_1996} & 26 &  \cite{mahmood2019improving,bertrand_role_2015,whitlock2018interacting,zhang_flowmatic_2020,ricca2020influence,endo_modularhmd_2021,malinov_stuckinspace_2021,amaro2022design,kim2023active,chen2023leap,liu2024visioncoach,Garcia2024speaking,Song2024Exploring,ahmmed2024system,liu2024visioncoach,cauquis2024investigating,zou2024effect,Acevedo2024effects,wang2024a3rt,monty2024analysis,Teo2019Technique,Safikhani2021Influence,Shi2022Groupbased,Ullah2023Exploring,Holly2024Gamebased,Kim2024Lifter} \\ \midrule
\textbf{Perception} & & \textit{\textbf{7}} & \\
& IOS \cite{gachter_measuring_2015} & 4 & \cite{dey_effects_2017, harris_asymmetry_2019,chen2024effects,cheng2024first} \\
& VEQ \cite{Roth_2020_veq} & 3 &\cite{Landeck2023From,merz2024voice,merz2024universal} \\
\bottomrule
\end{tabularx}
\Description{Summary tables of questionnaires used to evaluate user experience in virtual environments. Most common are SSQ for sickness, NASA-TLX for task load, PQ/IPQ for presence, SAM for emotion, and SUS for usability. Other dimensions include engagement, experience, immersion, and perception with fewer questionnaires.}
\end{table*}

\begin{figure*}[ht]
    \centering
    \includegraphics[width=\textwidth]{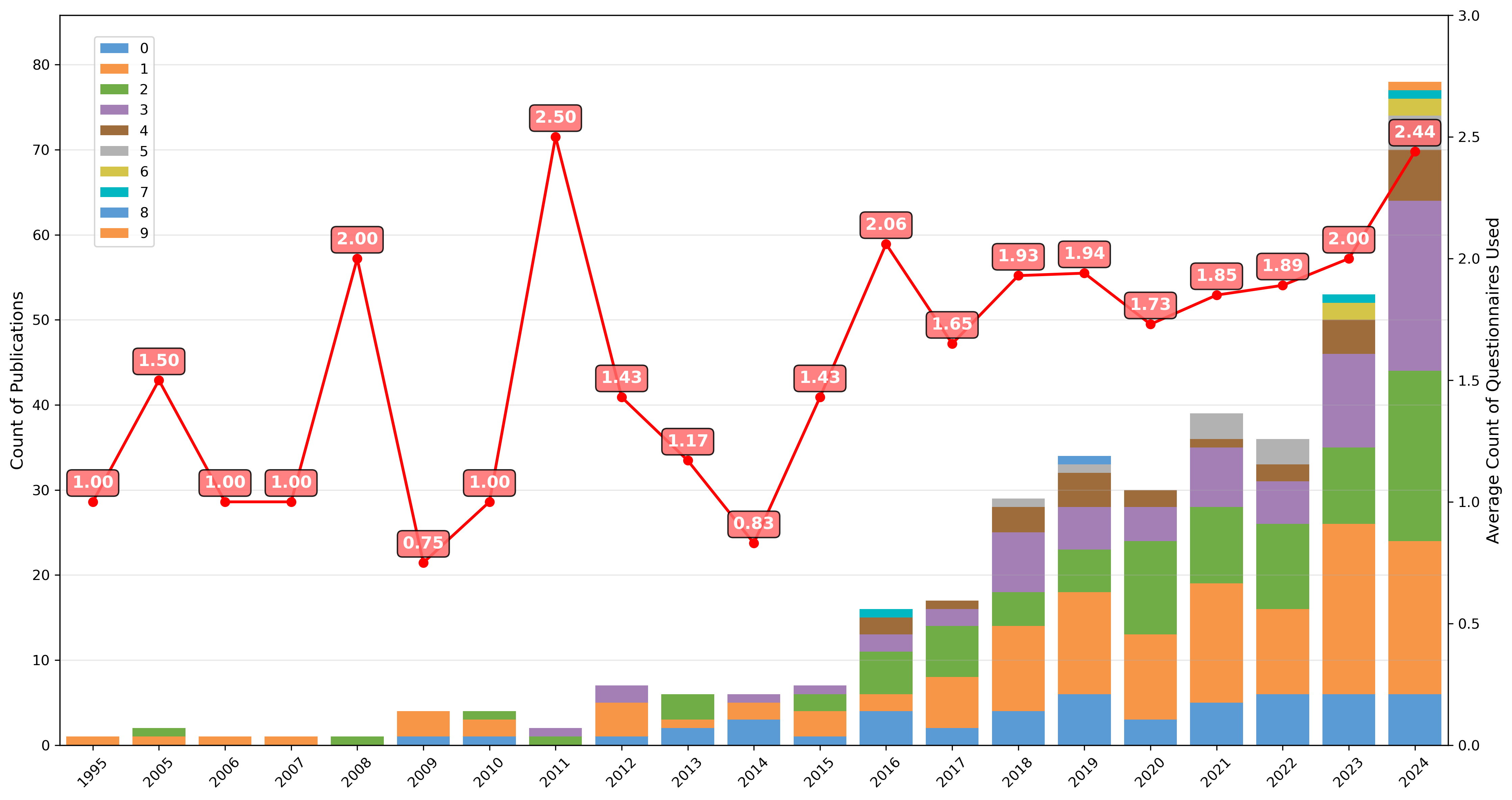}
    \caption{%
        A stacked bar chart illustrating the yearly distribution of publications using different numbers of questionnaires (represented by color-coded categories). 
        The red dashed line with circular markers represents the yearly weighted average count of questionnaires used. 
        The left y-axis corresponds to the count of publications, while the right y-axis corresponds to the average number of questionnaires used.
    }
    \Description{%
        A stacked bar chart showing the number of publications per year, grouped by the number of questionnaires used. 
        The x-axis represents years from 1995 to 2024, the left y-axis represents the count of publications, and the right y-axis represents the weighted average number of questionnaires used. 
        The bars are color-coded to indicate different questionnaire usage counts, and a red dashed line with circular markers represents the yearly weighted average.
    }
    \label{fig:publication_year}
\end{figure*}

We examined how many questionnaires are typically employed in research. As shown in \autoref{fig:publication_year}, early work (prior to 2010) almost always relied on a single questionnaire, whereas more recent publications tend to combine multiple standardized questionnaires within the same evaluation. The red dashed line indicates that the yearly weighted average number of questionnaires per paper has gradually risen over time, reaching values close to 2.44 in recent years. We note, however, that the elevated averages observed in some intermediate years (e.g., the peak of 2.50 in 2011) are largely artifacts of our ICs/ECs, which produced relatively small publication counts in those years. These fluctuations therefore do not reflect stable practice. A more conservative interpretation suggests that the typical number of questionnaires used prior to 2015 was likely closer to around 1.0–1.5 overall, despite occasional spikes. The long-term upward trajectory, rather than isolated yearly peaks, represents the more reliable pattern.

From 2017 onward, this upward trend becomes substantially clearer. As can be seen from the taller green and higher-order stacked segments in \autoref{fig:publication_year}, a growing proportion of studies employ three or more standardized questionnaires within a single evaluation. Cases involving four or five questionnaires---and in some instances as many as eight, the maximum observed in our dataset---shift from being rare exceptions to recurring occurrences between 2019 and 2024. This marks a distinct methodological departure from pre-2010 work, where single-questionnaire evaluations dominated, and from the 2010–2015 period, where multi-questionnaire usage appeared only sporadically. While earlier peaks must be interpreted with caution due to limited sample sizes, the consistent post-2017 increase, supported by both rising averages and substantially larger publication volumes, indicates a broader field-wide move toward richer, multi-instrument evaluation strategies.

\subsection{Observing Behavioral Patterns}

We marked a paper as reporting \emph{Task Performance} when it included objective measures of participants' performance on experimental tasks. These measures go beyond subjective ratings and capture observable behaviors, such as task completion time, accuracy, error rate, number of attempts, path efficiency, and many others. They often reveal how immersive environments alter motor performance, attention, and decision-making across different technical and contextual conditions.

Across the results, task performance frequently complemented questionnaires by providing behavioral evidence about how users responded to immersive systems. For example,~\citet{lee_exploring_2017} analyzed completion time and number of attempts in an AR ``pit'' task, showing that participants approached the task more cautiously and required more attempts when confronted with a visually threatening drop, even though frame-rate degradations alone had little effect. This illustrates how environmental design can shape users’ behavioral responses in ways that reflect their experiential state. In redirection research, \citet{bolte2015subliminal} used two-alternative forced choice tasks to estimate thresholds at which scene rotations and translations during saccades became detectable. Rather than merely reporting accuracy, these thresholds defined the limits within which spatial manipulations remained perceptually unnoticed, offering a direct behavioral grounding for when presence is likely to remain intact. In haptic realism studies, such as reported by~\citet{cai2020thermairglove}, measures of material-identification accuracy, decision time, and workload demonstrated that thermal cues supported reliable object discrimination without adding effort, suggesting improved perceptual fidelity rather than simply faster or slower task execution. Productivity-oriented evaluations, such as in VR text entry~\cite{grubert2018text}, used metrics like words per minute, error rate, initiation latency, and head posture to assess whether everyday tasks remained viable in immersive settings. Here, stable accuracy and manageable posture adjustments indicated that efficient performance could be maintained. Locomotion studies such as the one reported by~\citet{freitag_examining_2016} incorporated search times, spatial-memory scores, dual-task reaction times, and rotation behavior to examine how users adapted to rotation gain in CAVE environments, showing that performance often remained robust despite substantial perceptual manipulations. Taken together, these examples demonstrate that performance metrics serve not only as indicators of task success but also as behavioral probes into how system properties, sensory manipulations, and environmental cues shape users’ perceptual and experiential states.

These examples illustrate the breadth of how task performance is operationalized. Researchers variously used accuracy, speed, error rate, decision thresholds, workload, and behavioral adaptations as objective markers of how immersion shaped user activity. Importantly, such measures consistently linked immersion to observable performance changes: threatening environments slowed performance, subliminal manipulations preserved continuity, haptic cues improved object recognition, text entry retained usability in VR, and locomotion gains prompted adaptive movement strategies. These behavioral outcomes provide converging evidence that immersion is not only felt but enacted through measurable patterns in how people move, decide, and perform within virtual environments.

\subsection{Tracking System Performance}

In our coding scheme, we marked papers as reporting System Performance when they provided objective measurements of a system’s technical capabilities that were largely independent of direct user interaction. Such metrics included latency, frame rate, actuation strength, tracking accuracy, rendering stability, computational or power load, and related properties. These measures reflect how well a system functions technically, and often serve either as manipulation checks or as primary evidence for immersion-related claims.

Across the dataset, researchers reported a wide range of system-level measurements. For instance, several haptic systems quantified the \emph{fidelity and strength of actuator outputs}. Peiris et al.~\cite{peiris_thermovr_2017} assessed recognition accuracy and response times for thermal cues in a head-mounted display, while Chen et al.~\cite{Chen_haptivec_2019} measured solenoid output force, power draw, and directional recognition accuracy in a tactile pin-array controller. Fan et al.~\cite{fan2024spinshot} benchmarked torque, impulse duration, and latency in a flywheel-based impact device, demonstrating technical improvements over prior ungrounded haptics. These examples highlight how hardware-oriented work often validates immersion claims through quantitative evidence that signals are strong, separable, and delivered with low delay.

Beyond haptics, some studies focused on \emph{audio, rendering, and networking quality}. For example, Kao et al.~\cite{kao_audio_2022} controlled for audial system fidelity by normalizing voice recordings, validating their distinctiveness, and profiling WebGL rendering stability to ensure that observed effects on immersion were not confounded by performance bottlenecks. Likewise, telepresence research such as Higuchi et al.~\cite{higuchi_immerseboard_2015} directly measured gaze-estimation accuracy, gesture-tracking response times, and reconstruction fidelity, linking system accuracy to the sense of co-presence during remote collaboration.

Together, these examples illustrate a broader pattern. When researchers explicitly report system performance, they typically focus on metrics such as signal fidelity, actuation capacity, timing stability, and rendering robustness. Importantly, these measures are not only technical baselines but are often interpreted as prerequisites for immersive experience, showing that immersion depends on the system being accurate, responsive, and reliable at a technical level.

\subsection{Capturing Physiological Signals}

Physiological signals were occasionally employed as complementary indicators of immersion, providing an embodied layer of evidence beyond self-report. Researchers drew on a wide spectrum of biosignals, for example, heart rate and heart rate variability, electrodermal activity (EDA/GSR), respiration, EEG, and pupil size, to probe stress, arousal, relaxation, or cybersickness during immersive tasks.

Some studies focused on relaxation and affective comfort. For instance, Ito et al.~\cite{Ito_2023_wind} monitored ECG, respiration, and EEG while participants experienced simulated wind in VR, finding that heart rate variability indices reflected a measurable relaxation effect even when respiration and EEG did not change significantly. Likewise, Kim et al.~\cite{kim2023active} tracked heart rate, HRV, and oxygen saturation with consumer wearables, showing that active interventions such as VR drawing or throwing promoted stress recovery more effectively than passive viewing.

Physiological data have also been used to capture arousal and fear responses. In aviation safety training, Chittaro et al.~\cite{Chittaro2015assessing} reported stronger electrodermal and cardiovascular activation in immersive VR than in traditional safety cards, linking heightened arousal to improved long-term knowledge retention. Work on cybersickness provides another angle: Jung et al.~\cite{jung_floor-vibration_2021} showed that adding synchronized floor vibration significantly reduced skin conductance responses during VR driving, indicating lower stress and improved comfort. Early explorations in AR~\cite{gandy2010experiences} also investigated heart rate, skin conductance, and temperature in a ``pit'' scenario, though results were noisy and highlighted the sensitivity of these measures to contextual and procedural factors.

Across these examples, physiological measures provided converging evidence that immersive systems can influence bodily states, sometimes reinforcing subjective reports of relaxation or stress, other times exposing subtle effects not captured by questionnaires. At the same time, their applicability is limited to specific settings, as biosignals are highly sensitive to movement, task demands, and environmental context. These constraints restrict when physiological measures can be meaningfully deployed and underline the methodological care required to interpret them.

\subsection{Engaging in Interviews and Discussions}

Interviews and post-use discussions were employed in many studies ($N=169$, 45.1\%) to capture aspects of immersion that structured measures could not easily address. These exchanges often targeted themes such as enjoyment, agency, realism, comfort, and social connectedness---dimensions that reflect how participants themselves made sense of their experiences. For example, in stress recovery studies~\cite{kim2023active}, comments about creative freedom, cathartic release, or fatigue explained why active VR interventions felt more engaging and stress-relieving than passive viewing. Participants described the haptic beams as ``natural'' and ``fun,'' yet pointed out breaks in immersion caused by device weight or latency~\cite{ryu_elastick_2020}, highlighting how ergonomics and fidelity jointly shaped presence. Interviewees contrasted anticipatory and onset air-jet cues. Anticipatory cues provided predictability that reduced sickness, whereas onset cues felt more realistic due to their tight temporal synchrony~\cite{ke_turnahead_2023}. When anticipatory timing was perceived as implausible, it was reported as immersion-breaking, underscoring the importance of temporal fit. Even brief debriefs indicated that unnoticed scene edits helped preserve perceptual continuity and user comfort, lending support to the core mechanism of subliminal redirection~\cite{bolte2015subliminal}.

Overall, interviews and open-ended discussions with participants served as a flexible lens for revealing why immersion strengthened or weakened. They highlighted that presence is reinforced when users feel creative agency, sensorimotor congruence, or predictable feedback, and weakened when fidelity gaps, discomfort, or communication barriers intrude. In this way, qualitative reflections grounded the numerical data, offering nuanced explanations of how immersive systems succeed or fail from the user’s perspective.

\section{When to Use What Measure(s)?}

Our analysis reveals that the choice of evaluation method is rarely arbitrary. Instead, it reflects the priorities of our coding schemes, i.e., devices, primary contribution types, and particular domains. Below, we summarize recurring patterns observed across the 375 reviewed papers.

\paragraph{By Device.} 
Patterns also emerge at the device level. 
VR studies overwhelmingly employ SSQ, IPQ, and task performance, consistent with both presence research and simulator sickness concerns. 
AR and MR studies, however, emphasize pragmatic workload (e.g., NASA--TLX) and usability, reflecting their integration into real-world contexts. 
CAVE-based work often relies on basic task and system metrics, with relatively little use of presence questionnaires. 
Other devices (e.g., custom headsets or hybrid setups) demonstrate more heterogeneous strategies but still lean toward performance and system logging.

\paragraph{By Contribution Type.} 
Similarly, evaluation choices align with the type of contribution. 
Application papers tend to use straightforward task and interview measures. 
Design papers privilege qualitative feedback, with interviews dominating alongside SSQ or workload scales. 
Novel interaction techniques are almost always validated through task comparisons, often supplemented by presence or usability questionnaires. 
System evaluations place stronger emphasis on system-level logging and fidelity metrics, while usability-focused studies combine task efficiency, user satisfaction, and workload measures.

\paragraph{By Application Domain.} 
In gaming and entertainment, task performance is almost always paired with engagement-oriented questionnaires such as GEQ or IEQ, reflecting the emphasis on enjoyment. 
Health and rehabilitation studies, by contrast, prioritize safety and comfort, with SSQ and physiological measures far more common than in other domains. 
Navigation and locomotion research consistently combines task completion with simulator sickness metrics, while perception and sensory feedback studies foreground psychophysical tasks and system-level fidelity. 
Social interaction and collaboration papers rely heavily on interviews to capture nuances of communication, presence, and co-experience. 
Training and education contexts frequently combine performance outcomes with pragmatic usability scales, while UI/UX and design evaluations almost uniformly adopt a triad of task metrics, usability questionnaires, and workload measures.

\section{Discussion, Limitations, and Future Work}

\begin{figure*}[t]
    \centering
    \includegraphics[width=\linewidth]{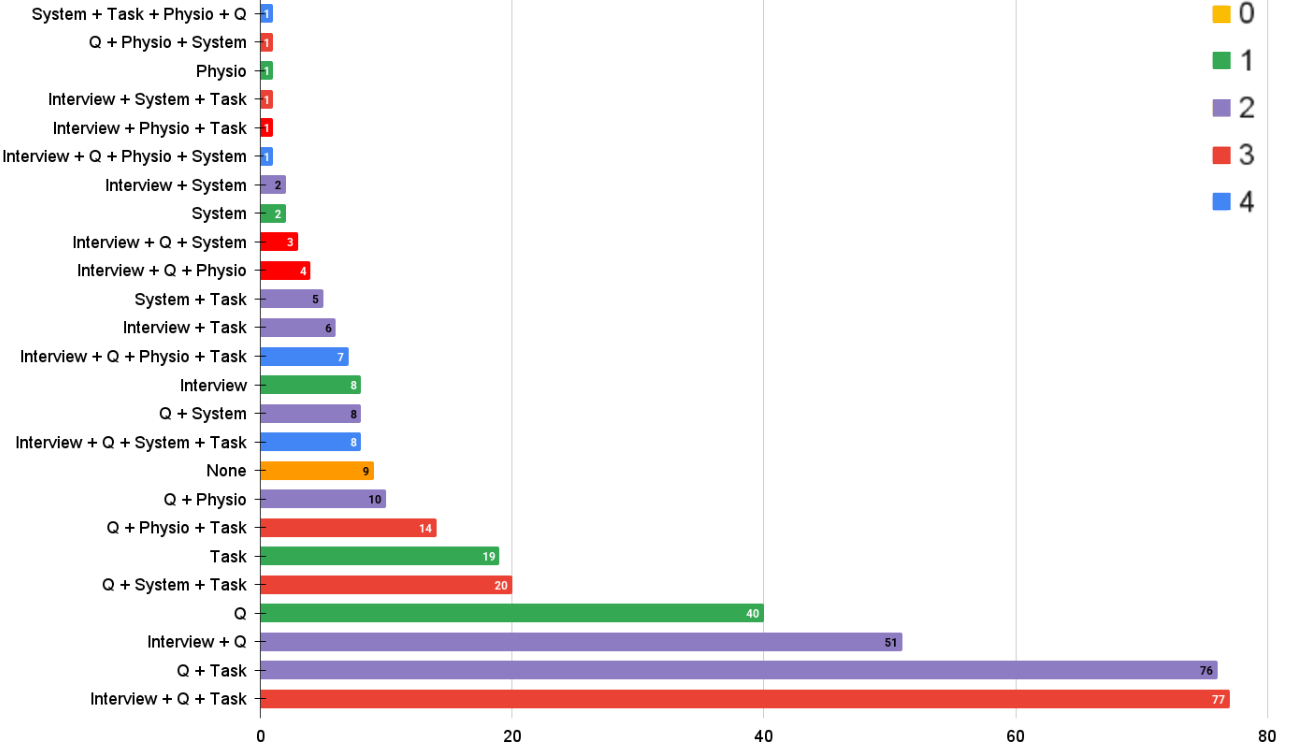}
    \caption{Distribution of immersive experience evaluation method combinations across 375 reviewed papers. Bars represent the number of papers using each specific combination of evaluation methods. Colors indicate the number of distinct evaluation method types employed in each case, ranging from zero to four. The most common combinations involve questionnaires and behavioral measures, with fewer studies integrating physiological or system-level metrics.}
    \Description{A horizontal bar chart showing the frequency of different combinations of UX evaluation methods used in immersive systems papers. Each bar is labeled with a combination such as ``Q + Task + Interview'' or ``Q only.'' Bars are color-coded by the number of methods used: gray for zero, blue for one, orange for two, red for three, and purple for four. The most frequent combination is ``Q + Task + Interview,'' followed by ``Q + Task” and “Q + Interview.'' Few papers used four methods, and only one used physiological + interview + system + questionnaire.}
    \label{fig:evaluation_distribution}
\end{figure*}

\subsection{How Do We \emph{Really} Evaluate Immersive Experiences?}  

Our findings make clear that immersive experience cannot be captured by a single, universal recipe. Instead, evaluation practices emerge as deeply entangled with domain priorities and technical affordances. The point is not merely that ``different domains use different tools.'' Rather, these patterns illustrate how constructs are selectively foregrounded or sidelined depending on context. What counts as ``experience'' in one community—fun, flow, enjoyment~\cite{nacke2008flow}---may be irrelevant in another where safety or fidelity~\cite{al2022framework} is the overriding concern. This domain-sensitivity helps explain why efforts to impose a universal evaluation framework have repeatedly failed. At the same time, our data show a striking imbalance in method adoption. Questionnaires and task metrics dominate, often combined with interviews, while more resource-intensive approaches such as physiological sensing or system-level logging remain rare. This asymmetry is not accidental: questionnaires are easy to administer and compare, whereas physiological and system metrics require equipment, expertise, and longer pipelines. The result is a methodological skew toward what is feasible rather than what is always conceptually appropriate.  

Taken together, these observations suggest that immersive experience evaluation is shaped as much by disciplinary priorities and resource constraints as by theoretical considerations (see \autoref{fig:evaluation_distribution}). Recognizing this helps us move beyond the search for a one-size-fits-all solution and instead ask: how can we design evaluations that are sensitive to context, but also cumulative enough to build a coherent knowledge base?

\subsection{Why Is the Field Still So Fragmented?}  
Despite decades of progress, immersive experience evaluation remains highly fragmented. Core constructs, such as presence, continue to be defined and operationalized in multiple, sometimes conflicting ways~\cite{slater_2009,skarbez_2017}. Similarly, embodiment has been carefully decomposed into agency, self-location, and body ownership~\cite{Kilteni_2012}, yet measurement practices vary widely, with some studies relying on custom questionnaires and others drawing on indirect behavioral proxies~\cite{peck_2024}. Such inconsistencies limit comparability across studies and make cumulative knowledge difficult to achieve.  

Methodological inconsistencies further exacerbate this issue. Even well-established instruments such as PQ, IPQ, and NASA-TLX are often used in modified or partial forms, sometimes without proper citation to their original sources. We observed examples where reviews were cited instead of the validated scales, or where additional items were introduced without validation. As Messick~\cite{messick1989validity,messick1995validity} emphasizes, validity is not a property of the measure itself but of the meaning of the scores and the inferences drawn from them. This highlights a broader problem: it is not the mere frequency of a measure’s use that guarantees its validity, but whether its underlying constructs align with the phenomena under study. Questionnaires carry assumptions that may not always fit the context, and similar concerns apply to other methods with respect to reliability~\cite{albert2022measuring}.

\subsection{It's Not Just About More Measures, It's About Smarter Measures}  

More is not necessarily better. We found more and more cases in recent years where six or more questionnaires were administered within a single study, creating redundancy and participant fatigue without yielding deeper insight. The relative infrequency of physiological measures should not be mistaken for neglect; rather, such measures are only appropriate in specific settings, as they require controlled movement, stable sensing conditions, and tasks that do not confound biosignals. When these conditions are met, they are commonly used and to good effect. By contrast, system-level performance metrics were often reported only superficially, making it difficult to relate technical properties to experiential outcomes. Taken together, these practices leave the field with a patchwork of methods that lack coherence.

One clear trend in our dataset is the proliferation of measures per study. Researchers increasingly adopt multi-method evaluations, combining questionnaires, task performance, and interviews. Yet this does not automatically resolve conceptual ambiguity. More measures do not guarantee better evaluation. Administering multiple questionnaires that probe overlapping constructs may only increase redundancy, while overwhelming participants and diluting insights.

The key is smarter combinations. For example, pairing a presence questionnaire with a task-based metric can provide complementary insights into both subjective experience and behavioral outcomes. In contrast, adding five additional questionnaires about presence-like constructs may not add conceptual clarity. Multi-method evaluation should not be treated as a checklist but as a principled process of selecting methods that jointly address the constructs of interest. This requires both theoretical reflection on what immersive experience means in context and careful methodological design to ensure integration rather than accumulation.

\subsection{The Missing Piece: User-Centric Evaluation}  
A striking absence in much of the literature is direct attention to what users themselves value and experience. Many evaluations remain researcher-centric, focusing on constructs that are easy to measure rather than those most salient to participants. Related work has shown the importance of dimensions such as workload~\cite{kosch_2023}, empathy~\cite{lee_2024}, and embodiment~\cite{Kilteni_2012,peck_2024}, yet these aspects are often sidelined in favor of traditional presence or usability scales. Our review shows that qualitative methods such as interviews are increasingly adopted, but still underrepresented compared to questionnaires. 

We therefore argue that immersive experience evaluation should shift toward more user-centric approaches. Interviews and open-ended reports can surface unexpected aspects of experience that structured scales may miss, such as social connectedness in collaborative XR or affective resonance in embodied interfaces. Future work should also pay attention to the missing aspects of immersive experience, for example, emotions, values, and lived contexts that are not well captured by existing instruments. Without such perspectives, evaluations risk reducing experience to a narrow set of technical constructs.

However, as with user testing more broadly, such qualitative and user‑centric approaches should not be expected to yield dramatic or surprising insights in every instance~\cite{lieberman2003tyranny}. Their true value lies in the consistent, iterative uncovering of both small frictions and occasional blind spots, which over time can lead to more holistic and meaningful understandings of immersive experience.

\subsection{Can Computational Modeling Help?}

Computational modeling can complement traditional evaluation by providing predictive and objective insights. The NICER metric, for example, models endurance and recovery in mid-air gestures and correlates more strongly with subjective fatigue than prior cumulative-fatigue models~\cite{yi2022nicer}. SIM2VR takes a different approach, integrating biomechanical simulation into VR applications to predict user performance and ergonomics without requiring large-scale user studies~\cite{fischer2024sim2vr}. Singh et al.~\cite{singh2024eevr} present EEVR, a dataset pairing physiological signals with textual emotion reports for joint representation learning. While designed for affective computing, EEVR illustrates the potential of cross-modal strategies that integrate physiological, behavioral, and linguistic data into more interpretable models of experience. Although these examples demonstrate the promise of computational approaches, they do not yet constitute a general or systematic methodology for modeling immersive experience.

The value of these approaches becomes most evident when combined with human-centered methods. NICER and its extension in AlphaPig \cite{li2025alphapig} show that computational metrics gain validity only when aligned with self-reports and task outcomes, while SIM2VR requires calibration against real user data. Long-term and socially situated XR use underscores this need for hybrid approaches. Biener et al.~\cite{biener2022quantifying} demonstrated that working in VR for a week produces fatigue and usability issues beyond what post-experience questionnaires capture, and Pavanatto et al.~\cite{pavanatto2025working} revealed how social acceptability and bystander reactions shape XR use in public settings. In such contexts, computational models can provide continuous traces of fatigue, workload, or social dynamics, while qualitative reports explain whether these signals genuinely reflect lived experience. Together, these examples suggest that hybrid approaches, which combine predictive modeling with user-centric insights, offer one of the most promising paths for future evaluation of immersive experiences.

\subsection{Looking Ahead: Towards an Open and Sustainable Evaluation Ecosystem}  
Beyond methods, the field must also consider infrastructures for sustaining progress. Our review found that few studies share protocols or datasets, limiting reproducibility \cite{feick2020virtual,borhani2024enhancing}. This gap persists despite emerging efforts to support reproducible immersive research. For instance, the Virtual Experience Research Accelerator (VERA\footnote{\url{https://sreal.ucf.edu/vera/}}) aims to provide a scalable, standardized environment for XR experimentation. Such initiatives show that shared infrastructure is feasible and increasingly valued, yet adoption remains uneven and fragmented.

We envision an open platform (similar to Locomotion Vault~\cite{di2021locomotion}) where researchers can share evaluation protocols, including details about domains, devices, and measures. Such a platform would serve as a living resource, evolving with new technologies and supporting cross-project comparability. To be effective, it should follow FAIR principles\footnote{\url{https://www.go-fair.org/fair-principles/}} by making protocols and datasets findable via persistent identifiers, accessible under clear licensing, interoperable across XR toolchains, and reusable through versioned documentation. This would enable replication, multi-lab comparison, and reuse without repeated re-implementation. Alongside this, repositories of validated scales and multimodal datasets could provide foundations for both traditional and computational approaches. Ultimately, we imagine a dynamic field guide for immersive experience evaluation: a community-maintained resource that enables newcomers, researchers, and practitioners to identify context-specific best practices while building cumulative knowledge. 

\subsection{Limitations}

Our review has several limitations that should be considered when interpreting the findings. First, our dataset is not comprehensive. It does not cover all immersive-technology venues and therefore inevitably reflects the methodological preferences and biases of the technical ACM/IEEE communities from which it is drawn. In line with similar scoping efforts~\cite{speicher2019mixed,bergstrom2021evaluate}, this constitutes a common constraint of reviews that target specific publication ecosystems rather than the entire interdisciplinary landscape. However, given that our methodology intentionally adopts a bottom-up, venue-specific structure, this paper does not aim to provide a fully comprehensive literature review of all immersive technologies involving user experience. Instead, it seeks to synthesize representative evaluation metrics and methodological patterns from seven premier HCI/XR venues. Despite these boundaries, we believe our review still offers meaningful guidance for understanding current evaluation practices and for supporting researchers and practitioners when designing and assessing novel interaction techniques in specific application contexts.

Second, our keyword-based retrieval strategy has inherent constraints. While the selection of ``immersive experience'' terms was iteratively refined to balance precision and recall, keyword search cannot capture all relevant work, particularly when terminology diverges across communities or when commonly expected descriptors (e.g., ``CAVE''), are intentionally avoided due to trademark issues. Future work could complement this approach with citation-based, snowball, or embedding-based retrieval methods to minimize such blind spots.

Third, although we employed a multi-stage screening and coding process, coding consistency and methodological decisions introduce potential bias. Short papers were excluded to ensure full reporting of study design and measurement details, yet this may underrepresent application-centered or situated evaluations that often appear as short formats at ACM CHI or other venues. We relied on dual independent coding with consensus resolution instead of formal inter-rater reliability statistics, which prioritizes interpretative agreement but limits claims of measurement reliability. To support transparency and mitigate these concerns, we release our full dataset and PRISMA documentation, enabling others to inspect our decisions and replicate or extend the analysis.

Finally, our synthesis and recommendations are shaped by the above boundaries. The gaps we highlight, such as fragmented terminology, limited cross-study comparability, or the need for shared repositories, should be interpreted as gaps within the selected venues, not universal deficits across all XR communities. In parallel, other fields, for example, digital heritage~\cite{wang2025grand,pujol2019did,pujol2012evaluating}, have developed their own evaluation traditions and frameworks that were beyond the scope of this review. In 3D heritage, Galeazzi and Di Giuseppantonio Di Franco~\cite{galeazzi2017theorising} emphasize that evaluation in immersive archaeology must be integrated and cognitively informed, covering perception, presence, and usability. Koutsabasis~\cite{koutsabasis2017empirical} further synthesizes empirical methods and validity challenges for interactive cultural heritage systems. A more cross-disciplinary synthesis that integrates these communities, alongside emerging infrastructures such as VERA or questionnaire toolkits, represents a valuable direction for future work.

\section{In Closing}

The evaluation of immersive experience will never be one-size-fits-all. Best practice is to remain context-sensitive~\cite{greenberg2008usability}, methodologically transparent, and user-centered. At the same time, the community must confront conceptual fragmentation and methodological inconsistency, while also embracing emerging opportunities in computational modeling and open science. Moving forward requires collective effort: clarifying constructs, adopting open protocols, and developing integrative frameworks. We see this paper as one step toward such a future, and we invite researchers and practitioners to reflect on their own evaluation choices and to contribute to building a more transparent, cumulative, and user-focused ecosystem for immersive experience research.

Against this backdrop, our scoping review of 375 papers across ACM CHI, UIST, VRST, SUI, IEEE VR, ISMAR, and TVCG revealed a field rich in methods but fractured in coherence. Immersive experience has been assessed through questionnaires, task and system performance, interviews, and physiological signals, yet no single approach dominates across contexts, and even well-established constructs are applied in inconsistent ways. From this fragmented landscape, three insights stand out. First, evaluation is domain-sensitive. For example, what matters in gaming is not the same as in health, education, or collaboration. Second, more measures do not automatically mean better evaluation; the challenge is to design smarter, conceptually aligned combinations. Third, progress requires infrastructure: open protocols, shared datasets, and hybrid approaches that integrate computational modeling with user-centered insights. 

We do not claim to resolve decades of debate, nor to prescribe a universal framework. Instead, our contribution is to consolidate existing practices, expose the assumptions that shape them, and highlight where gaps remain. Lasting progress will depend on collective engagement to strengthen methodological transparency and to shape evaluation practices that are not only rigorous but also sustainable.

\section{Data Availability}
All data generated or analyzed during this study are included in its supplementary information files.


\begin{acks}
Xiang Li is supported by the China Scholarship Council (CSC) International Cambridge Scholarship (No. 202208320092). Per Ola Kristensson is supported by the EPSRC (grant EP/W02456X/1). We only used a Large Language Model (ChatGPT) to assist with grammar checking and language polishing. All scientific decisions (screening process, analysis, discussions, and conclusions) are entirely the responsibility of the authors. The use of ChatGPT did not influence, manipulate, or generate substantive content for this research.
\end{acks}

\bibliographystyle{ACM-Reference-Format}
\bibliography{sample-base}

@article{pujol2019did,
  title={Did we just travel to the past? Building and evaluating with cultural presence different modes of VR-mediated experiences in virtual archaeology},
  author={Pujol-Tost, Laia},
  journal={Journal on Computing and Cultural Heritage (JOCCH)},
  volume={12},
  number={1},
  pages={1--20},
  year={2019},
  publisher={ACM New York, NY, USA}
}

@article{KNOWLES2020101928,
title = {Specificity of trait anxiety in anxiety and depression: Meta-analysis of the State-Trait Anxiety Inventory},
journal = {Clinical Psychology Review},
volume = {82},
pages = {101928},
year = {2020},
issn = {0272-7358},
doi = {https://doi.org/10.1016/j.cpr.2020.101928},
url = {https://www.sciencedirect.com/science/article/pii/S0272735820301161},
author = {Kelly A. Knowles and Bunmi O. Olatunji}
}

@article{nowak_2003_presence,
    author = {Nowak, Kristine L. and Biocca, Frank},
    title = {The Effect of the Agency and Anthropomorphism on Users' Sense of Telepresence, Copresence, and Social Presence in Virtual Environments},
    journal = {Presence: Teleoperators and Virtual Environments},
    volume = {12},
    number = {5},
    pages = {481-494},
    year = {2003},
    month = {10},
    doi = {10.1162/105474603322761289},
    url = {https://doi.org/10.1162/105474603322761289},
}

@article{jeremy_2003_spatial,
author = {Jeremy N. Bailenson and Jim Blascovich and Andrew C. Beall and Jack M. Loomis},
title ={Interpersonal Distance in Immersive Virtual Environments},

journal = {Personality and Social Psychology Bulletin},
volume = {29},
number = {7},
pages = {819-833},
year = {2003},
doi = {10.1177/0146167203029007002},
    note ={PMID: 15018671},
URL = { 
        https://doi.org/10.1177/0146167203029007002   
},
eprint = { 
        https://doi.org/10.1177/0146167203029007002
}
}

@ARTICLE{Roth_2020_veq,
  author={Roth, Daniel and Latoschik, Marc Erich},
  journal={IEEE Transactions on Visualization and Computer Graphics}, 
  title={Construction of the Virtual Embodiment Questionnaire (VEQ)}, 
  year={2020},
  volume={26},
  number={12},
  pages={3546-3556},
  doi={10.1109/TVCG.2020.3023603}}

@article{Lessiter2001ITC,
    author = {Lessiter, Jane and Freeman, Jonathan and Keogh, Edmund and Davidoff, Jules},
    title = {A Cross-Media Presence Questionnaire: The ITC-Sense of Presence Inventory},
    journal = {Presence: Teleoperators and Virtual Environments},
    volume = {10},
    number = {3},
    pages = {282-297},
    year = {2001},
    month = {06},
    doi = {10.1162/105474601300343612},
    url = {https://doi.org/10.1162/105474601300343612},
    eprint = {https://direct.mit.edu/pvar/article-pdf/10/3/282/1623678/105474601300343612.pdf},
}

@book{Spielberger1970STAI,
  title        = {STAI: Manual for the State-Trait Anxiety Inventory},
  author       = {Spielberger, Charles D. and Gorsuch, Richard L. and Lushene, Robert E.},
  year         = {1970},
  publisher    = {Consulting Psychologists Press},
  address      = {Palo Alto}
}

@book{Spielberger_1983_STAI,
author = {Spielberger, Charles and Gorsuch, Richard and Lushene, Robert and Vagg, PR and Jacobs, Gerard},
year = {1983},
month = {01},
pages = {},
title = {Manual for the State-Trait Anxiety Inventory (Form Y1 – Y2)},
volume = {IV},
journal = {Palo Alto, CA: Consulting Psychologists Press;}
}

@article{li2024investigating,
  title={Investigating creation perspectives and icon placement preferences for on-body menus in virtual reality},
  author={Li, Xiang and He, Wei and Jin, Shan and Gugenheimer, Jan and Hui, Pan and Liang, Hai-Ning and Kristensson, Per Ola},
  journal={Proceedings of the ACM on Human-Computer Interaction},
  volume={8},
  number={ISS},
  pages={236--254},
  year={2024},
  publisher={ACM New York, NY, USA}
}

@article{wang2025grand,
  title={Grand challenges in immersive technologies for cultural heritage},
  author={Wang, Hanbing and Du, Junyan and Li, Yue and Zhang, Lie and Li, Xiang},
  journal={International Journal of Human--Computer Interaction},
  pages={1--22},
  year={2025},
  publisher={Taylor \& Francis}
}

@ARTICLE{li2025evaluating,
  author={Li, Xiang and He, Wei and Kristensson, Per Ola},
  journal={IEEE Transactions on Visualization and Computer Graphics}, 
  title={Evaluating the Usability of Microgestures for Text Editing Tasks in Virtual Reality}, 
  year={2025},
  volume={},
  number={},
  pages={1-14},
  doi={10.1109/TVCG.2025.3642050}}

@inproceedings{xiao2025concept,
  title={A Concept at Work: A Review of Motivations, Operationalizations, and Conclusions in VR Research about Presence},
  author={Xiao, Cleo and Yu, Difeng and Hornb{\ae}k, Kasper and Bergstr{\"o}m, Joanna},
  booktitle={Proceedings of the 2025 CHI Conference on Human Factors in Computing Systems},
  pages={1--21},
  year={2025}
}

@inproceedings{li2025bend,
  title={Bend It, Aim It, Tap It: Designing an On-Body Disambiguation Mechanism for Curve Selection in Mixed Reality},
  author={Li, Xiang and Kristensson, Per Ola},
  booktitle={Proceedings of the 2025 ACM Symposium on Spatial User Interaction},
  pages={1--12},
  year={2025}
}

@article{wang2025handows,
  title={Handows: A Palm-Based Interactive Multi-Window Management System in Virtual Reality},
  author={Wang, Jin-Du and Zhou, Ke and Ren, Haoyu and Kristensson, Per Ola and Li, Xiang},
  journal={IEEE Transactions on Visualization and Computer Graphics},
  year={2025},
  publisher={IEEE}
}

@inproceedings{Ariza2016Inducing,
author = {Ariza, Oscar and Freiwald, Jann and Laage, Nadine and Feist, Michaela and Salloum, Mariam and Bruder, Gerd and Steinicke, Frank},
title = {Inducing Body-Transfer Illusions in VR by Providing Brief Phases of Visual-Tactile Stimulation},
year = {2016},
isbn = {9781450340687},
publisher = {Association for Computing Machinery},
address = {New York, NY, USA},
url = {https://doi.org/10.1145/2983310.2985760},
doi = {10.1145/2983310.2985760},
booktitle = {Proceedings of the 2016 Symposium on Spatial User Interaction},
pages = {61–68},
numpages = {8},
location = {Tokyo, Japan},
series = {SUI '16}
}

@inproceedings{Bozgeyikli2016Locomotion,
author = {Bozgeyikli, Evren and Raij, Andrew and Katkoori, Srinivas and Dubey, Rajiv},
title = {Locomotion in Virtual Reality for Individuals with Autism Spectrum Disorder},
year = {2016},
isbn = {9781450340687},
publisher = {Association for Computing Machinery},
address = {New York, NY, USA},
url = {https://doi.org/10.1145/2983310.2985763},
doi = {10.1145/2983310.2985763},
booktitle = {Proceedings of the 2016 Symposium on Spatial User Interaction},
pages = {33–42},
numpages = {10},
location = {Tokyo, Japan},
series = {SUI '16}
}

@inproceedings{Valkov2017Smooth,
author = {Valkov, Dimitar and Flagge, Steffen},
title = {Smooth immersion: the benefits of making the transition to virtual environments a continuous process},
year = {2017},
isbn = {9781450354868},
publisher = {Association for Computing Machinery},
address = {New York, NY, USA},
url = {https://doi.org/10.1145/3131277.3132183},
doi = {10.1145/3131277.3132183},
booktitle = {Proceedings of the 5th Symposium on Spatial User Interaction},
pages = {12–19},
numpages = {8},
location = {Brighton, United Kingdom},
series = {SUI '17}
}

@inproceedings{Lee2021Design,
author = {Lee, Chia-Yang and Hsieh, Wei-An and Brickler, David and Babu, Sabarish V. and Chuang, Jung-Hong},
title = {Design and Empirical Evaluation of a Novel Near-field Interaction Metaphor on Distant Object Manipulation in VR},
year = {2021},
isbn = {9781450390910},
publisher = {Association for Computing Machinery},
address = {New York, NY, USA},
url = {https://doi.org/10.1145/3485279.3485296},
doi = {10.1145/3485279.3485296},
booktitle = {Proceedings of the 2021 ACM Symposium on Spatial User Interaction},
articleno = {13},
numpages = {11},
location = {Virtual Event, USA},
series = {SUI '21}
}

@inproceedings{Shi2022Groupbased,
author = {Shi, Rongkai and Zhang, Jialin and Stuerzlinger, Wolfgang and Liang, Hai-Ning},
title = {Group-based Object Alignment in Virtual Reality Environments},
year = {2022},
isbn = {9781450399487},
publisher = {Association for Computing Machinery},
address = {New York, NY, USA},
url = {https://doi.org/10.1145/3565970.3567682},
doi = {10.1145/3565970.3567682},
booktitle = {Proceedings of the 2022 ACM Symposium on Spatial User Interaction},
articleno = {2},
numpages = {11},
location = {Online, CA, USA},
series = {SUI '22}
}

@inproceedings{Podkosova2023Joint,
author = {Podkosova, Iana and De Pace, Francesco and Brument, Hugo},
title = {Joint Action in Collaborative Mixed Reality: Effects of Immersion Type and Physical Location},
year = {2023},
isbn = {9798400702815},
publisher = {Association for Computing Machinery},
address = {New York, NY, USA},
url = {https://doi.org/10.1145/3607822.3614541},
doi = {10.1145/3607822.3614541},
booktitle = {Proceedings of the 2023 ACM Symposium on Spatial User Interaction},
articleno = {16},
numpages = {12},
location = {Sydney, NSW, Australia},
series = {SUI '23}
}

@inproceedings{Zhang2023Forestlight,
author = {Zhang, Jingyu and Jiang, Ke and Wang, Suhan and Ming, Shijie and Wang, Huidi},
title = {Forestlight: A Virtual Reality Respiratory Biofeedback System Using Interactive Lighting for Pressure Relief},
year = {2023},
isbn = {9798400702815},
publisher = {Association for Computing Machinery},
address = {New York, NY, USA},
url = {https://doi.org/10.1145/3607822.3616414},
doi = {10.1145/3607822.3616414},
booktitle = {Proceedings of the 2023 ACM Symposium on Spatial User Interaction},
articleno = {64},
numpages = {5},
location = {Sydney, NSW, Australia},
series = {SUI '23}
}

@inproceedings{Ilo2024Goldilocks,
author = {Ilo, Cory and DiVerdi, Stephen and Bowman, Doug},
title = {Goldilocks Zoning: Evaluating a Gaze-Aware Approach to Task-Agnostic VR Notification Placement},
year = {2024},
isbn = {9798400710889},
publisher = {Association for Computing Machinery},
address = {New York, NY, USA},
url = {https://doi.org/10.1145/3677386.3682087},
doi = {10.1145/3677386.3682087},
booktitle = {Proceedings of the 2024 ACM Symposium on Spatial User Interaction},
articleno = {13},
numpages = {12},
location = {Trier, Germany},
series = {SUI '24}
}

@inproceedings{Cools2024Impact,
author = {Cools, Robbe and Venema, Ren\'{e}e and Esteves, Augusto and Simeone, Adalberto L.},
title = {The Impact of Near-Future Mixed Reality Contact Lenses on Users' Lives via an Immersive Speculative Enactment and Focus Groups},
year = {2024},
isbn = {9798400710889},
publisher = {Association for Computing Machinery},
address = {New York, NY, USA},
url = {https://doi.org/10.1145/3677386.3682099},
doi = {10.1145/3677386.3682099},
booktitle = {Proceedings of the 2024 ACM Symposium on Spatial User Interaction},
articleno = {9},
numpages = {13},
location = {Trier, Germany},
series = {SUI '24}
}

@inproceedings{Zielinski2013Intercept,
author = {Zielinski, David J. and Kopper, Regis and McMahan, Ryan P. and Lu, Wenjie and Ferrari, Silvia},
title = {Intercept tags: enhancing intercept-based systems},
year = {2013},
isbn = {9781450323796},
publisher = {Association for Computing Machinery},
address = {New York, NY, USA},
url = {https://doi.org/10.1145/2503713.2503737},
doi = {10.1145/2503713.2503737},
booktitle = {Proceedings of the 19th ACM Symposium on Virtual Reality Software and Technology},
pages = {263–266},
numpages = {4},
location = {Singapore},
series = {VRST '13}
}

@inproceedings{Chen20136dof,
author = {Chen, Weiya and Plancoulaine, Anthony and F\'{e}rey, Nicolas and Touraine, Damien and Nelson, Julien and Bourdot, Patrick},
title = {6DoF navigation in virtual worlds: comparison of joystick-based and head-controlled paradigms},
year = {2013},
isbn = {9781450323796},
publisher = {Association for Computing Machinery},
address = {New York, NY, USA},
url = {https://doi.org/10.1145/2503713.2503754},
doi = {10.1145/2503713.2503754},
booktitle = {Proceedings of the 19th ACM Symposium on Virtual Reality Software and Technology},
pages = {111–114},
numpages = {4},
location = {Singapore},
series = {VRST '13}
}

@inproceedings{Llorach2014Simulator,
author = {Llorach, Gerard and Evans, Alun and Blat, Josep},
title = {Simulator sickness and presence using HMDs: comparing use of a game controller and a position estimation system},
year = {2014},
isbn = {9781450332538},
publisher = {Association for Computing Machinery},
address = {New York, NY, USA},
url = {https://doi.org/10.1145/2671015.2671120},
doi = {10.1145/2671015.2671120},
booktitle = {Proceedings of the 20th ACM Symposium on Virtual Reality Software and Technology},
pages = {137–140},
numpages = {4},
location = {Edinburgh, Scotland},
series = {VRST '14}
}

@inproceedings{Arafat2016Effects,
author = {Arafat, Imtiaz Muhammad and Ferdous, Sharif Mohammad Shahnewaz and Quarles, John},
title = {The effects of cybersickness on persons with multiple sclerosis},
year = {2016},
isbn = {9781450344913},
publisher = {Association for Computing Machinery},
address = {New York, NY, USA},
url = {https://doi.org/10.1145/2993369.2993383},
doi = {10.1145/2993369.2993383},
booktitle = {Proceedings of the 22nd ACM Conference on Virtual Reality Software and Technology},
pages = {51–59},
numpages = {9},
location = {Munich, Germany},
series = {VRST '16}
}

@inproceedings{Westhoven2016Head,
author = {Westhoven, Martin and Paul, Dennis and Alexander, Thomas},
title = {Head turn scaling below the threshold of perception in immersive virtual environments},
year = {2016},
isbn = {9781450344913},
publisher = {Association for Computing Machinery},
address = {New York, NY, USA},
url = {https://doi.org/10.1145/2993369.2993385},
doi = {10.1145/2993369.2993385},
booktitle = {Proceedings of the 22nd ACM Conference on Virtual Reality Software and Technology},
pages = {77–86},
numpages = {10},
location = {Munich, Germany},
series = {VRST '16}
}

@inproceedings{Argelaguet2016Giant,
author = {Argelaguet, Ferran and Maignant, Morgant},
title = {GiAnt: stereoscopic-compliant multi-scale navigation in VEs},
year = {2016},
isbn = {9781450344913},
publisher = {Association for Computing Machinery},
address = {New York, NY, USA},
url = {https://doi.org/10.1145/2993369.2993391},
doi = {10.1145/2993369.2993391},
booktitle = {Proceedings of the 22nd ACM Conference on Virtual Reality Software and Technology},
pages = {269–277},
numpages = {9},
location = {Munich, Germany},
series = {VRST '16}
}

@inproceedings{Chowdhury2017Information,
author = {Chowdhury, Tanvir Irfan and Ferdous, Sharif Mohammad Shahnewaz and Quarles, John},
title = {Information recall in a virtual reality disability simulation},
year = {2017},
isbn = {9781450355483},
publisher = {Association for Computing Machinery},
address = {New York, NY, USA},
url = {https://doi.org/10.1145/3139131.3139143},
doi = {10.1145/3139131.3139143},
booktitle = {Proceedings of the 23rd ACM Symposium on Virtual Reality Software and Technology},
articleno = {37},
numpages = {10},
location = {Gothenburg, Sweden},
series = {VRST '17}
}

@inproceedings{Skarbez2018Immersion,
author = {Skarbez, Richard and Brooks, Frederick P. and Whitton, Mary C.},
title = {Immersion and coherence in a stressful virtual environment},
year = {2018},
isbn = {9781450360869},
publisher = {Association for Computing Machinery},
address = {New York, NY, USA},
url = {https://doi.org/10.1145/3281505.3281530},
doi = {10.1145/3281505.3281530},
booktitle = {Proceedings of the 24th ACM Symposium on Virtual Reality Software and Technology},
articleno = {24},
numpages = {11},
location = {Tokyo, Japan},
series = {VRST '18}
}

@inproceedings{Teo2019Technique,
author = {Teo, Theophilus and F. Hayati, Ashkan and A. Lee, Gun and Billinghurst, Mark and Adcock, Matt},
title = {A Technique for Mixed Reality Remote Collaboration using 360 Panoramas in 3D Reconstructed Scenes},
year = {2019},
isbn = {9781450370011},
publisher = {Association for Computing Machinery},
address = {New York, NY, USA},
url = {https://doi.org/10.1145/3359996.3364238},
doi = {10.1145/3359996.3364238},
booktitle = {Proceedings of the 25th ACM Symposium on Virtual Reality Software and Technology},
articleno = {23},
numpages = {11},
location = {Parramatta, NSW, Australia},
series = {VRST '19}
}

@inproceedings{Min2020Effects,
author = {Min, Seulki and Moon, Jung-geun and Cho, Chul-Hyun and Kim, Gerard J.},
title = {Effects of Immersive Virtual Reality Content Type to Mindfulness and Physiological Parameters},
year = {2020},
isbn = {9781450376198},
publisher = {Association for Computing Machinery},
address = {New York, NY, USA},
url = {https://doi.org/10.1145/3385956.3418942},
doi = {10.1145/3385956.3418942},
booktitle = {Proceedings of the 26th ACM Symposium on Virtual Reality Software and Technology},
articleno = {24},
numpages = {9},
location = {Virtual Event, Canada},
series = {VRST '20}
}

@inproceedings{Safikhani2021Influence,
author = {Safikhani, Saeed and Holly, Michael and Kainz, Alexander and Pirker, Johanna},
title = {The Influence of in-VR Questionnaire Design on the User Experience},
year = {2021},
isbn = {9781450390927},
publisher = {Association for Computing Machinery},
address = {New York, NY, USA},
url = {https://doi.org/10.1145/3489849.3489884},
doi = {10.1145/3489849.3489884},
booktitle = {Proceedings of the 27th ACM Symposium on Virtual Reality Software and Technology},
articleno = {12},
numpages = {8},
location = {Osaka, Japan},
series = {VRST '21}
}

@inproceedings{Gauthier2021Virtual,
author = {Gauthier, Baptiste and Albert, Louis and Martuzzi, Roberto and Herbelin, Bruno and Blanke, Olaf},
title = {Virtual Reality platform for functional magnetic resonance imaging in ecologically valid conditions},
year = {2021},
isbn = {9781450390927},
publisher = {Association for Computing Machinery},
address = {New York, NY, USA},
url = {https://doi.org/10.1145/3489849.3489894},
doi = {10.1145/3489849.3489894},
booktitle = {Proceedings of the 27th ACM Symposium on Virtual Reality Software and Technology},
articleno = {36},
numpages = {12},
location = {Osaka, Japan},
series = {VRST '21}
}

@inproceedings{Ratcliffe2022Rich,
author = {Ratcliffe, Jack and Tokarchuk, Laurissa},
title = {Rich virtual feedback from sensorimotor interaction may harm, not help, learning in immersive virtual reality},
year = {2022},
isbn = {9781450398893},
publisher = {Association for Computing Machinery},
address = {New York, NY, USA},
url = {https://doi.org/10.1145/3562939.3565633},
doi = {10.1145/3562939.3565633},
booktitle = {Proceedings of the 28th ACM Symposium on Virtual Reality Software and Technology},
articleno = {14},
numpages = {12},
location = {Tsukuba, Japan},
series = {VRST '22}
}

@inproceedings{Somarathna2023Exploring,
author = {Somarathna, Rukshani and Elvitigala, Don Samitha and Yan, Yijun and Quigley, Aaron J and Mohammadi, Gelareh},
title = {Exploring User Engagement in Immersive Virtual Reality Games through Multimodal Body Movements},
year = {2023},
isbn = {9798400703287},
publisher = {Association for Computing Machinery},
address = {New York, NY, USA},
url = {https://doi.org/10.1145/3611659.3615687},
doi = {10.1145/3611659.3615687},
booktitle = {Proceedings of the 29th ACM Symposium on Virtual Reality Software and Technology},
articleno = {3},
numpages = {8},
location = {Christchurch, New Zealand},
series = {VRST '23}
}

@inproceedings{Jung2023Crossreality,
author = {Jung, Sungchul and Wu, Yuanjie and Lukosch, Stephan and Lukosch, Heide and Mckee, Ryan Douglas and Lindeman, Robert W.},
title = {Cross-Reality Gaming: Comparing Competition and Collaboration in an Asymmetric Gaming Experience},
year = {2023},
isbn = {9798400703287},
publisher = {Association for Computing Machinery},
address = {New York, NY, USA},
url = {https://doi.org/10.1145/3611659.3615698},
doi = {10.1145/3611659.3615698},
booktitle = {Proceedings of the 29th ACM Symposium on Virtual Reality Software and Technology},
articleno = {4},
numpages = {10},
location = {Christchurch, New Zealand},
series = {VRST '23}
}

@inproceedings{Landeck2023From,
author = {Landeck, Maximilian and Unruh, Fabian and Lugrin, Jean-Luc and Latoschik, Marc Erich},
title = {From Clocks to Pendulums: A Study on the Influence of External Moving Objects on Time Perception in Virtual Environments},
year = {2023},
isbn = {9798400703287},
publisher = {Association for Computing Machinery},
address = {New York, NY, USA},
url = {https://doi.org/10.1145/3611659.3615703},
doi = {10.1145/3611659.3615703},
booktitle = {Proceedings of the 29th ACM Symposium on Virtual Reality Software and Technology},
articleno = {15},
numpages = {11},
location = {Christchurch, New Zealand},
series = {VRST '23}
}

@inproceedings{Ullah2023Exploring,
author = {Ullah, A K M Amanat and Delamare, William and Hasan, Khalad},
title = {Exploring Users' Pointing Performance on Virtual and Physical Large Curved Displays},
year = {2023},
isbn = {9798400703287},
publisher = {Association for Computing Machinery},
address = {New York, NY, USA},
url = {https://doi.org/10.1145/3611659.3615710},
doi = {10.1145/3611659.3615710},
booktitle = {Proceedings of the 29th ACM Symposium on Virtual Reality Software and Technology},
articleno = {7},
numpages = {11},
location = {Christchurch, New Zealand},
series = {VRST '23}
}

@inproceedings{Gabel2023Redirecting,
author = {Gabel, Jenny and Schmidt, Susanne and Ariza, Oscar and Steinicke, Frank},
title = {Redirecting Rays: Evaluation of Assistive Raycasting Techniques in Virtual Reality},
year = {2023},
isbn = {9798400703287},
publisher = {Association for Computing Machinery},
address = {New York, NY, USA},
url = {https://doi.org/10.1145/3611659.3615716},
doi = {10.1145/3611659.3615716},
booktitle = {Proceedings of the 29th ACM Symposium on Virtual Reality Software and Technology},
articleno = {38},
numpages = {11},
location = {Christchurch, New Zealand},
series = {VRST '23}
}

@inproceedings{Christiansen2024Exploring,
author = {Christiansen, Frederik Roland and Hollensberg, Linus N\o{}rgaard and Jensen, Niko Bach and Julsgaard, Kristian and Jespersen, Kristian Nyborg and Nikolov, Ivan},
title = {Exploring Presence in Interactions with LLM-Driven NPCs: A Comparative Study of Speech Recognition and Dialogue Options},
year = {2024},
isbn = {9798400705359},
publisher = {Association for Computing Machinery},
address = {New York, NY, USA},
url = {https://doi.org/10.1145/3641825.3687716},
doi = {10.1145/3641825.3687716},
booktitle = {Proceedings of the 30th ACM Symposium on Virtual Reality Software and Technology},
articleno = {6},
numpages = {11},
location = {Trier, Germany},
series = {VRST '24}
}

@inproceedings{Xu2024Istraypaws,
author = {Xu, Yao and Ding, Ding and Chen, Yongxin and Li, Zhuying and Xu, Xiangyu},
title = {iStrayPaws: Immersing in a Stray Animal's World through First-Person VR to Bridge Human-Animal Empathy},
year = {2024},
isbn = {9798400705359},
publisher = {Association for Computing Machinery},
address = {New York, NY, USA},
url = {https://doi.org/10.1145/3641825.3687729},
doi = {10.1145/3641825.3687729},
booktitle = {Proceedings of the 30th ACM Symposium on Virtual Reality Software and Technology},
articleno = {5},
numpages = {11},
location = {Trier, Germany},
series = {VRST '24}
}

@inproceedings{Holly2024Gamebased,
author = {Holly, Michael and Brettschuh, Sandra and Tiwari, Ajay Shankar and Bhagat, Kaushal Kumar and Pirker, Johanna},
title = {Game-Based Motivation: Enhancing Learning with Achievements in a Customizable Virtual Reality Environment},
year = {2024},
isbn = {9798400705359},
publisher = {Association for Computing Machinery},
address = {New York, NY, USA},
url = {https://doi.org/10.1145/3641825.3687741},
doi = {10.1145/3641825.3687741},
booktitle = {Proceedings of the 30th ACM Symposium on Virtual Reality Software and Technology},
articleno = {30},
numpages = {11},
location = {Trier, Germany},
series = {VRST '24}
}

@inproceedings{Eroglu2024Choose,
author = {Eroglu, Sevinc and Schmitz, Patric and Sinke, Kilian and Anders, David and Kuhlen, Torsten Wolfgang and Weyers, Benjamin},
title = {Choose Your Reference Frame Right: An Immersive Authoring Technique for Creating Reactive Behavior},
year = {2024},
isbn = {9798400705359},
publisher = {Association for Computing Machinery},
address = {New York, NY, USA},
url = {https://doi.org/10.1145/3641825.3687744},
doi = {10.1145/3641825.3687744},
booktitle = {Proceedings of the 30th ACM Symposium on Virtual Reality Software and Technology},
articleno = {17},
numpages = {11},
location = {Trier, Germany},
series = {VRST '24}
}

@inproceedings{Kim2024Lifter,
author = {Kim, JaeHoon and Joo, DongYun and Shin, Hyemin and Lee, Sun-Uk and Kim, Gerard Jounghyun and Kim, Hanseob},
title = {Lifter for VR Headset: Enhancing Immersion, Presence, Flow, and Alleviating Mental and Physical Fatigue during Prolonged Use},
year = {2024},
isbn = {9798400705359},
publisher = {Association for Computing Machinery},
address = {New York, NY, USA},
url = {https://doi.org/10.1145/3641825.3687753},
doi = {10.1145/3641825.3687753},
booktitle = {Proceedings of the 30th ACM Symposium on Virtual Reality Software and Technology},
articleno = {2},
numpages = {12},
location = {Trier, Germany},
series = {VRST '24}
}

@misc{IxDF2023Immersion,
  author       = {{Interaction Design Foundation - IxDF}},
  title        = {What is Immersion in Extended Reality (XR)?},
  year         = {2023},
  howpublished = {\url{https://www.interaction-design.org/literature/topics/immersion}},
  note         = {Retrieved November 18, 2025}
}

@article{chen2024being,
  title={What is ‘being there’? An ontology of the immersive experience},
  author={Chen, Chen and Hu, Xiaohan and Fisher, Jacob},
  journal={Annals of the International Communication Association},
  volume={48},
  number={4},
  pages={391--414},
  year={2024},
  publisher={Oxford University Press}
}

@inproceedings{feick2020virtual,
  title={The virtual reality questionnaire toolkit},
  author={Feick, Martin and Kleer, Niko and Tang, Anthony and Kr{\"u}ger, Antonio},
  booktitle={Adjunct proceedings of the 33rd annual ACM symposium on user interface software and technology},
  pages={68--69},
  year={2020}
}

@article{regenbrecht2002measuring,
  title={Measuring presence in augmented reality environments: design and a first test of a questionnaire},
  author={Regenbrecht, Holger and Schubert, Thomas},
  journal={arXiv preprint arXiv:2103.02831},
  year={2002}
}

@book{hale2014handbook,
  title={Handbook of virtual environments: Design, implementation, and applications},
  author={Hale, Kelly S and Stanney, Kay M},
  year={2014},
  publisher={CRC Press}
}

@inproceedings{jerald2017human,
  title={Human-centered design for immersive interactions},
  author={Jerald, Jason},
  booktitle={2017 IEEE Virtual Reality (VR)},
  pages={431--432},
  year={2017},
  organization={IEEE}
}

@book{laviola20173d,
  title={3D user interfaces: theory and practice},
  author={LaViola Jr, Joseph J and Kruijff, Ernst and McMahan, Ryan P and Bowman, Doug and Poupyrev, Ivan P},
  year={2017},
  publisher={Addison-Wesley Professional}
}

@inproceedings{bergstrom2021evaluate,
  title={How to evaluate object selection and manipulation in vr? guidelines from 20 years of studies},
  author={Bergstr{\"o}m, Joanna and Dalsgaard, Tor-Salve and Alexander, Jason and Hornb{\ae}k, Kasper},
  booktitle={proceedings of the 2021 CHI conference on human factors in computing systems},
  pages={1--20},
  year={2021}
}

@inproceedings{speicher2019mixed,
  title={What is mixed reality?},
  author={Speicher, Maximilian and Hall, Brian D and Nebeling, Michael},
  booktitle={Proceedings of the 2019 CHI conference on human factors in computing systems},
  pages={1--15},
  year={2019}
}

@inproceedings{borhani2024enhancing,
  title={Enhancing replicability in XR HCI studies: a survey-based approach},
  author={Borhani, Zahra and Ortega, Francisco R},
  booktitle={2024 IEEE International Symposium on Mixed and Augmented Reality Adjunct (ISMAR-Adjunct)},
  pages={42--46},
  year={2024},
  organization={IEEE}
}

@article{pujol2012evaluating,
  title={Evaluating presence in cultural heritage projects},
  author={Pujol, Laia and Champion, Erik},
  journal={International Journal of Heritage Studies},
  volume={18},
  number={1},
  pages={83--102},
  year={2012},
  publisher={Taylor \& Francis}
}

@article{galeazzi2017theorising,
  title={Theorising 3D Visualisation Systems in Archaeology: Towards more effective design, evaluations and life cycles},
  author={Galeazzi, Fabrizio and Di Giuseppantonio Di Franco, Paola},
  year={2017},
  publisher={Council for British Archaeology}
}

@article{koutsabasis2017empirical,
  title={Empirical evaluations of interactive systems in cultural heritage: A review},
  author={Koutsabasis, Panayiotis},
  journal={International Journal of Computational Methods in Heritage Science (IJCMHS)},
  volume={1},
  number={1},
  pages={100--122},
  year={2017},
  publisher={IGI Global Scientific Publishing}
}

@article{yi2022nicer,
  title={Nicer: A new and improved consumed endurance and recovery metric to quantify muscle fatigue of mid-air interactions},
  author={Li, Yi and Tag, Benjamin and Dai, Shaozhang and Crowther, Robert and Dwyer, Tim and Irani, Pourang and Ens, Barrett},
  journal={ACM Transactions on Graphics (TOG)},
  volume={43},
  number={4},
  pages={1--14},
  year={2024},
  publisher={ACM New York, NY, USA}
}

@article{lieberman2003tyranny,
  title={The tyranny of evaluation},
  author={Lieberman, Henry},
  journal={Retrieved October},
  volume={2},
  pages={2008},
  year={2003}
}

@inproceedings{nacke2008flow,
  title={Flow and immersion in first-person shooters: measuring the player's gameplay experience},
  author={Nacke, Lennart and Lindley, Craig A},
  booktitle={Proceedings of the 2008 conference on future play: Research, play, share},
  pages={81--88},
  year={2008}
}

@article{singh2024eevr,
  title={EEVR: A Dataset of Paired Physiological Signals and Textual Descriptions for Joint Emotion Representation Learning},
  author={Singh, Pragya and Budhiraja, Ritvik and Gupta, Ankush and Goswami, Anshul and Kumar, Mohan and Singh, Pushpendra},
  journal={Advances in Neural Information Processing Systems},
  volume={37},
  pages={15765--15778},
  year={2024}
}

@article{al2022framework,
  title={A framework for fidelity evaluation of immersive virtual reality systems},
  author={Al-Jundi, Hamza A and Tanbour, Emad Y},
  journal={Virtual Reality},
  volume={26},
  number={3},
  pages={1103--1122},
  year={2022},
  publisher={Springer}
}

@article{biener2022quantifying,
  title={Quantifying the effects of working in VR for one week},
  author={Biener, Verena and Kalamkar, Snehanjali and Nouri, Negar and Ofek, Eyal and Pahud, Michel and Dudley, John J and Hu, Jinghui and Kristensson, Per Ola and Weerasinghe, Maheshya and Pucihar, Klen {\v{C}}opi{\v{c}} and others},
  journal={IEEE Transactions on Visualization and Computer Graphics},
  volume={28},
  number={11},
  pages={3810--3820},
  year={2022},
  publisher={IEEE}
}

@inproceedings{greenberg2008usability,
  title={Usability evaluation considered harmful (some of the time)},
  author={Greenberg, Saul and Buxton, Bill},
  booktitle={Proceedings of the SIGCHI conference on Human factors in computing systems},
  pages={111--120},
  year={2008}
}

@article{pavanatto2025working,
  title={Working in Extended Reality in the Wild: Worker and Bystander Experiences of XR Virtual Displays in Public Real-World Settings},
  author={Pavanatto, Leonardo and Biener, Verena and Chandran, Jennifer and Kalamkar, Snehanjali and Lu, Feiyu and Dudley, John J and Hu, Jinghui and Ramirez-Saffy, G Nikki and Kristensson, Per Ola and Giovannelli, Alexander and others},
  journal={IEEE Transactions on Visualization and Computer Graphics},
  year={2025},
  publisher={IEEE}
}

@inproceedings{li2025alphapig,
  title={Alphapig: The nicest way to prolong interactive gestures in extended reality},
  author={Li, Yi and Fischer, Florian and Dwyer, Tim and Ens, Barrett and Crowther, Robert and Kristensson, Per Ola and Tag, Benjamin},
  booktitle={Proceedings of the 2025 CHI Conference on Human Factors in Computing Systems},
  pages={1--14},
  year={2025}
}

@inproceedings{di2021locomotion,
  title={Locomotion vault: the extra mile in analyzing vr locomotion techniques},
  author={Di Luca, Massimiliano and Seifi, Hasti and Egan, Simon and Gonzalez-Franco, Mar},
  booktitle={Proceedings of the 2021 CHI conference on human factors in computing systems},
  pages={1--10},
  year={2021}
}

@incollection{messick1989validity,
  author       = {Messick, Samuel J.},
  title        = {Validity},
  editor       = {Linn, Robert L.},
  booktitle    = {Educational Measurement},
  edition      = {3rd},
  pages        = {13--103},
  year         = {1989},
  address      = {New York},
  publisher    = {Macmillan / American Council on Education}
}

@article{messick1995validity,
  author       = {Messick, Samuel J.},
  title        = {Validity of Psychological Assessment: Validation of Inferences from Persons’ Responses and Performances as Scientific Inquiry into Score Meaning},
  journal      = {American Psychologist},
  year         = {1995},
  volume       = {50},
  number       = {9},
  pages        = {741--749},
  doi          = {10.1037/0003-066X.50.9.741},
}

@inproceedings{fischer2024sim2vr,
  title={SIM2VR: Towards Automated Biomechanical Testing in VR},
  author={Fischer, Florian and Ikkala, Aleksi and Klar, Markus and Fleig, Arthur and Bachinski, Miroslav and Murray-Smith, Roderick and H{\"a}m{\"a}l{\"a}inen, Perttu and Oulasvirta, Antti and M{\"u}ller, J{\"o}rg},
  booktitle={Proceedings of the 37th Annual ACM Symposium on User Interface Software and Technology},
  pages={1--15},
  year={2024}
}

@book{albert2022measuring,
  title={Measuring the user experience: Collecting, analyzing, and presenting UX metrics},
  author={Albert, Bill and Tullis, Tom},
  year={2022},
  publisher={Morgan Kaufmann}
}

@INPROCEEDINGS{dey_2020,
  author={Dey, Arindam and Phoon, Jane and Saha, Shuvodeep and Dobbins, Chelsea and Billinghurst, Mark},
  booktitle={2020 IEEE International Symposium on Mixed and Augmented Reality (ISMAR)}, 
  title={A Neurophysiological Approach for Measuring Presence in Immersive Virtual Environments}, 
  year={2020},
  volume={},
  number={},
  pages={474-485},
  doi={10.1109/ISMAR50242.2020.00072}}

@article{Goncalves_2022,
author = {Gon\c{c}alves, Guilherme and Coelho, Hugo and Monteiro, Pedro and Melo, Miguel and Bessa, Maximino},
title = {Systematic Review of Comparative Studies of the Impact of Realism in Immersive Virtual Experiences},
year = {2022},
issue_date = {June 2023},
publisher = {Association for Computing Machinery},
address = {New York, NY, USA},
volume = {55},
number = {6},
issn = {0360-0300},
url = {https://doi.org/10.1145/3533377},
doi = {10.1145/3533377},
journal = {ACM Comput. Surv.},
month = dec,
articleno = {115},
numpages = {36}
}

@ARTICLE{zhang_2020,
  author={Zhang, Chenyan},
  journal={IEEE Access}, 
  title={The Why, What, and How of Immersive Experience}, 
  year={2020},
  volume={8},
  number={},
  pages={90878-90888},
  doi={10.1109/ACCESS.2020.2993646}}

@article{skarbez_2017,
author = {Skarbez, Richard and Brooks, Jr., Frederick P. and Whitton, Mary C.},
title = {A Survey of Presence and Related Concepts},
year = {2017},
issue_date = {November 2018},
publisher = {Association for Computing Machinery},
address = {New York, NY, USA},
volume = {50},
number = {6},
issn = {0360-0300},
url = {https://doi.org/10.1145/3134301},
doi = {10.1145/3134301},
journal = {ACM Comput. Surv.},
month = nov,
articleno = {96},
numpages = {39}
}

@article{souza_2021,
author = {Souza, Vinicius and Maciel, Anderson and Nedel, Luciana and Kopper, Regis},
title = {Measuring Presence in Virtual Environments: A Survey},
year = {2021},
issue_date = {November 2022},
publisher = {Association for Computing Machinery},
address = {New York, NY, USA},
volume = {54},
number = {8},
issn = {0360-0300},
url = {https://doi.org/10.1145/3466817},
doi = {10.1145/3466817},
journal = {ACM Comput. Surv.},
month = oct,
articleno = {163},
numpages = {37}
}

@article{cummings_2016,
  title={How immersive is enough? A meta-analysis of the effect of immersive technology on user presence},
  author={Cummings, James J and Bailenson, Jeremy N},
  journal={Media psychology},
  volume={19},
  number={2},
  pages={272--309},
  year={2016},
  publisher={Taylor \& Francis}
}

@ARTICLE{bowman_2007,
  author={Bowman, Doug A. and McMahan, Ryan P.},
  journal={Computer}, 
  title={Virtual Reality: How Much Immersion Is Enough?}, 
  year={2007},
  volume={40},
  number={7},
  pages={36-43},
  doi={10.1109/MC.2007.257}}

@ARTICLE{dey_2018,
  
AUTHOR={Dey, Arindam  and Billinghurst, Mark  and Lindeman, Robert W.  and Swan, J. Edward },
         
TITLE={A Systematic Review of 10 Years of Augmented Reality Usability Studies: 2005 to 2014},
        
JOURNAL={Frontiers in Robotics and AI},
        
VOLUME={Volume 5 - 2018},

YEAR={2018},

URL={https://www.frontiersin.org/journals/robotics-and-ai/articles/10.3389/frobt.2018.00037},

DOI={10.3389/frobt.2018.00037},

ISSN={2296-9144}}

@INPROCEEDINGS{zhou_2008,
  author={Feng Zhou and Duh, Henry Been-Lirn and Billinghurst, Mark},
  booktitle={2008 7th IEEE/ACM International Symposium on Mixed and Augmented Reality}, 
  title={Trends in augmented reality tracking, interaction and display: A review of ten years of ISMAR}, 
  year={2008},
  volume={},
  number={},
  pages={193-202},
  doi={10.1109/ISMAR.2008.4637362}}

@ARTICLE{kim_2018,
  author={Kim, Kangsoo and Billinghurst, Mark and Bruder, Gerd and Duh, Henry Been-Lirn and Welch, Gregory F.},
  journal={IEEE Transactions on Visualization and Computer Graphics}, 
  title={Revisiting Trends in Augmented Reality Research: A Review of the 2nd Decade of ISMAR (2008–2017)}, 
  year={2018},
  volume={24},
  number={11},
  pages={2947-2962},
  doi={10.1109/TVCG.2018.2868591}}

@inproceedings{dnser_2008,
  title={A survey of evaluation techniques used in augmented reality studies},
  author={Andreas D{\"u}nser and Rapha{\"e}l Grasset and Mark Billinghurst},
  booktitle={International Conference on Computer Graphics and Interactive Techniques},
  year={2008},
  url={https://api.semanticscholar.org/CorpusID:45775424}
}

@article{kim_2020,
author = {Yong Min Kim and Ilsun Rhiu and Myung Hwan Yun},
title = {A Systematic Review of a Virtual Reality System from the Perspective of User Experience},
journal = {International Journal of Human–Computer Interaction},
volume = {36},
number = {10},
pages = {893--910},
year = {2020},
publisher = {Taylor \& Francis},
doi = {10.1080/10447318.2019.1699746},
URL = { 
        https://doi.org/10.1080/10447318.2019.1699746
},
eprint = { 
        https://doi.org/10.1080/10447318.2019.1699746
}
}

@article{sanchez_2005,
  title={From presence to consciousness through virtual reality},
  author={Sanchez-Vives, Maria V and Slater, Mel},
  journal={Nature reviews neuroscience},
  volume={6},
  number={4},
  pages={332--339},
  year={2005},
  publisher={Nature Publishing Group UK London}
}

@article{spano_2023,
title = {Virtual nature, psychological and psychophysiological outcomes: A systematic review},
journal = {Journal of Environmental Psychology},
volume = {89},
pages = {102044},
year = {2023},
issn = {0272-4944},
doi = {https://doi.org/10.1016/j.jenvp.2023.102044},
url = {https://www.sciencedirect.com/science/article/pii/S0272494423000920},
author = {Giuseppina Spano and Annalisa Theodorou and Gerhard Reese and Giuseppe Carrus and Giovanni Sanesi and Angelo Panno}
}

@ARTICLE{wang_2023,
  author={Wang, Jialin and Shi, Rongkai and Zheng, Wenxuan and Xie, Weijie and Kao, Dominic and Liang, Hai-Ning},
  journal={IEEE Transactions on Visualization and Computer Graphics}, 
  title={Effect of Frame Rate on User Experience, Performance, and Simulator Sickness in Virtual Reality}, 
  year={2023},
  volume={29},
  number={5},
  pages={2478-2488},
  doi={10.1109/TVCG.2023.3247057}}

@ARTICLE{lee_2024,
  author={Lee, Yongho and Shin, Heesook and Gil, Youn-Hee},
  journal={IEEE Transactions on Visualization and Computer Graphics}, 
  title={Measurement of Empathy in Virtual Reality with Head-Mounted Displays: A Systematic Review}, 
  year={2024},
  volume={30},
  number={5},
  pages={2485-2495},
  doi={10.1109/TVCG.2024.3372076}}

@article{kosch_2023,
author = {Kosch, Thomas and Karolus, Jakob and Zagermann, Johannes and Reiterer, Harald and Schmidt, Albrecht and Wo\'{z}niak, Pawe\l{} W.},
title = {A Survey on Measuring Cognitive Workload in Human-Computer Interaction},
year = {2023},
issue_date = {December 2023},
publisher = {Association for Computing Machinery},
address = {New York, NY, USA},
volume = {55},
number = {13s},
issn = {0360-0300},
url = {https://doi.org/10.1145/3582272},
doi = {10.1145/3582272},
journal = {ACM Comput. Surv.},
month = jul,
articleno = {283},
numpages = {39}
}

@ARTICLE{peck_2024,
  author={Peck, Tabitha C. and Good, Jessica J.},
  journal={IEEE Transactions on Visualization and Computer Graphics}, 
  title={Measuring Embodiment: Movement Complexity and the Impact of Personal Characteristics}, 
  year={2024},
  volume={30},
  number={8},
  pages={4588-4600},
  doi={10.1109/TVCG.2023.3270725}}

@article{becker_2023,
author = {Becker, Artur and Freitas, Carla M. Dal Sasso},
title = {Evaluation of XR Applications: A Tertiary Review},
year = {2023},
issue_date = {May 2024},
publisher = {Association for Computing Machinery},
address = {New York, NY, USA},
volume = {56},
number = {5},
issn = {0360-0300},
url = {https://doi.org/10.1145/3626517},
doi = {10.1145/3626517},
journal = {ACM Comput. Surv.},
month = nov,
articleno = {110},
numpages = {35}
}

@article{felton_2022,
author = {William M. Felton and Russell E. Jackson},
title = {Presence: A Review},
journal = {International Journal of Human–Computer Interaction},
volume = {38},
number = {1},
pages = {1--18},
year = {2022},
publisher = {Taylor \& Francis},
doi = {10.1080/10447318.2021.1921368},
URL = { 
        https://doi.org/10.1080/10447318.2021.1921368
},
eprint = { 
        https://doi.org/10.1080/10447318.2021.1921368
    }
}

@article{nash_2000,
  title={A review of presence and performance in virtual environments},
  author={Nash, Eric B and Edwards, Gregory W and Thompson, Jennifer A and Barfield, Woodrow},
  journal={International Journal of human-computer Interaction},
  volume={12},
  number={1},
  pages={1--41},
  year={2000},
  publisher={Taylor \& Francis}
}

@INPROCEEDINGS{robert_2024,
  author={Robert, Florent and Wu, Hui-Yin and Sassatelli, Lucile and Winckler, Marco},
  booktitle={2024 IEEE Conference Virtual Reality and 3D User Interfaces (VR)}, 
  title={Task-based methodology to characterise immersive user experience with multivariate data}, 
  year={2024},
  volume={},
  number={},
  pages={722-731},
  doi={10.1109/VR58804.2024.00092}}

@article{bowman_1998,
  title={A methodology for the evaluation of travel techniques for immersive virtual environments},
  author={Bowman, Doug A and Koller, David and Hodges, Larry F},
  journal={Virtual reality},
  volume={3},
  number={2},
  pages={120--131},
  year={1998},
  publisher={Springer}
}

@inproceedings{McMahan_2006,
author = {McMahan, Ryan P. and Gorton, Doug and Gresock, Joe and McConnell, Will and Bowman, Doug A.},
title = {Separating the effects of level of immersion and 3D interaction techniques},
year = {2006},
isbn = {1595933212},
publisher = {Association for Computing Machinery},
address = {New York, NY, USA},
url = {https://doi.org/10.1145/1180495.1180518},
doi = {10.1145/1180495.1180518},
booktitle = {Proceedings of the ACM Symposium on Virtual Reality Software and Technology},
pages = {108–111},
numpages = {4},
location = {Limassol, Cyprus},
series = {VRST '06}
}

@article{slater_2003,
  title={A note on presence terminology},
  author={Slater, Mel},
  journal={Presence connect},
  volume={3},
  number={3},
  pages={1--5},
  year={2003}
}

@article{slater_2009,
  title={How we experience immersive virtual environments: the concept of presence and its measurement},
  author={Slater, Mel and Lotto, Beau and Arnold, Maria Marta and Sanchez-Vives, Maria V},
  journal={Anuario de psicolog{\'\i}a},
  volume={40},
  number={2},
  pages={193--210},
  year={2009},
  publisher={Universitat de Barcelona}
}

@article{slater_2010,
author = {Slater, Mel and Spanlang, Bernhard and Corominas, David},
title = {Simulating virtual environments within virtual environments as the basis for a psychophysics of presence},
year = {2010},
issue_date = {July 2010},
publisher = {Association for Computing Machinery},
address = {New York, NY, USA},
volume = {29},
number = {4},
issn = {0730-0301},
url = {https://doi.org/10.1145/1778765.1778829},
doi = {10.1145/1778765.1778829},
journal = {ACM Trans. Graph.},
month = jul,
articleno = {92},
numpages = {9}
}

@article{Kilteni_2012,
    author = {Kilteni, Konstantina and Groten, Raphaela and Slater, Mel},
    title = {The Sense of Embodiment in Virtual Reality},
    journal = {Presence: Teleoperators and Virtual Environments},
    volume = {21},
    number = {4},
    pages = {373-387},
    year = {2012},
    month = {11},
    doi = {10.1162/PRES_a_00124},
    url = {https://doi.org/10.1162/PRES_a_00124}
}

@article{slater_1994,
    author = {Slater, Mel and Usoh, Martin and Steed, Anthony},
    title = {Depth of Presence in Virtual Environments},
    journal = {Presence: Teleoperators and Virtual Environments},
    volume = {3},
    number = {2},
    pages = {130-144},
    year = {1994},
    month = {05},
    doi = {10.1162/pres.1994.3.2.130},
    url = {https://doi.org/10.1162/pres.1994.3.2.130},
    eprint = {https://direct.mit.edu/pvar/article-pdf/3/2/130/1622610/pres.1994.3.2.130.pdf},
}

@article{Murphy_2020,
    author = {Murphy, Dooley and Skarbez, Richard},
    title = {What Do We Mean When We Say ``Presence'''?},
    journal = {PRESENCE: Virtual and Augmented Reality},
    volume = {29},
    pages = {171-190},
    year = {2020},
    month = {12},
    issn = {1054-7460},
    doi = {10.1162/pres_a_00360},
    url = {https://doi.org/10.1162/pres_a_00360}
}

@article{tricco2018prisma,
  title={PRISMA extension for scoping reviews (PRISMA-ScR): checklist and explanation},
  author={Tricco, Andrea C and Lillie, Erin and Zarin, Wasifa and O'Brien, Kelly K and Colquhoun, Heather and Levac, Danielle and Moher, David and Peters, Micah DJ and Horsley, Tanya and Weeks, Laura and others},
  journal={Annals of internal medicine},
  volume={169},
  number={7},
  pages={467--473},
  year={2018},
  publisher={American College of Physicians}
}

@article{watson1988development,
  title={Development and validation of brief measures of positive and negative affect: the PANAS scales.},
  author={Watson, David and Clark, Lee Anna and Tellegen, Auke},
  journal={Journal of personality and social psychology},
  volume={54},
  number={6},
  pages={1063},
  year={1988},
  publisher={American Psychological Association}
}

@INPROCEEDINGS{Immohr2024evaluating,
  author={Immohr, Felix and Rendle, Gareth and Lammert, Anton and Neidhardt, Annika and Heyde, Victoria Meyer Zur and Froehlich, Bernd and Raake, Alexander},
  booktitle={2024 IEEE Conference Virtual Reality and 3D User Interfaces (VR)}, 
  title={Evaluating the Effect of Binaural Auralization on Audiovisual Plausibility and Communication Behavior in Virtual Reality}, 
  year={2024},
  volume={},
  number={},
  pages={849-858},
  doi={10.1109/VR58804.2024.00104}}

@INPROCEEDINGS{ju2016personality,
  author={Ju, Uijong and Kang, June and Wallraven, Christian},
  booktitle={2016 IEEE Virtual Reality (VR)}, 
  title={Personality differences predict decision-making in an accident situation in virtual driving}, 
  year={2016},
  volume={},
  number={},
  pages={77-82},
  doi={10.1109/VR.2016.7504690}}

@inproceedings{rogers_exploring_2019,
author = {Martin-Niedecken, Anna Lisa and Rogers, Katja and Turmo Vidal, Laia and Mekler, Elisa D. and M\'{a}rquez Segura, Elena},
title = {ExerCube vs. Personal Trainer: Evaluating a Holistic, Immersive, and Adaptive Fitness Game Setup},
year = {2019},
isbn = {9781450359702},
publisher = {Association for Computing Machinery},
address = {New York, NY, USA},
url = {https://doi.org/10.1145/3290605.3300318},
doi = {10.1145/3290605.3300318},
booktitle = {Proceedings of the 2019 CHI Conference on Human Factors in Computing Systems},
pages = {1–15},
numpages = {15},
location = {Glasgow, Scotland Uk},
series = {CHI '19}
}

@inproceedings{vanden2016design,
  title={Design and preliminary validation of the player experience inventory},
  author={Vanden Abeele, Vero and Nacke, Lennart E and Mekler, Elisa D and Johnson, Daniel},
  booktitle={Proceedings of the 2016 Annual Symposium on Computer-Human Interaction in Play Companion Extended Abstracts},
  pages={335--341},
  year={2016}
}

@inproceedings{xu_2019_pointing,
  author={Xu, Wenge and Liang, Hai-Ning and He, Anqi and Wang, Zifan},
  booktitle={2019 IEEE International Symposium on Mixed and Augmented Reality (ISMAR)}, 
  title={Pointing and Selection Methods for Text Entry in Augmented Reality Head Mounted Displays}, 
  year={2019},
  volume={},
  number={},
  pages={279-288},
  doi={10.1109/ISMAR.2019.00026}}

@INPROCEEDINGS{hwang_2023_enhancing,
  author={Hwang, Seokhyun and Kim, YoungIn and Seo, Youngseok and Kim, SeungJun},
  booktitle={2023 IEEE International Symposium on Mixed and Augmented Reality (ISMAR)}, 
  title={Enhancing Seamless Walking in Virtual Reality: Application of Bone-Conduction Vibration in Redirected Walking}, 
  year={2023},
  volume={},
  number={},
  pages={1181-1190},
  doi={10.1109/ISMAR59233.2023.00135}}

@book{Poels2007geq_book,
title = "D3.3 : Game Experience Questionnaire: development of a self-report measure to assess the psychological impact of digital games",
author = "K. Poels and {de Kort}, Y.A.W. and W.A. IJsselsteijn",
year = "2007",
language = "English",
publisher = "Technische Universiteit Eindhoven",
}

@article{BROCKMYER2009624,
title = {The development of the Game Engagement Questionnaire: A measure of engagement in video game-playing},
journal = {Journal of Experimental Social Psychology},
volume = {45},
number = {4},
pages = {624-634},
year = {2009},
issn = {0022-1031},
doi = {https://doi.org/10.1016/j.jesp.2009.02.016},
url = {https://www.sciencedirect.com/science/article/pii/S0022103109000444},
author = {Jeanne H. Brockmyer and Christine M. Fox and Kathleen A. Curtiss and Evan McBroom and Kimberly M. Burkhart and Jacquelyn N. Pidruzny}
}

@book{ijsselsteijn_game_2013,
	address = {Eindhoven},
	title = {The {Game} {Experience} {Questionnaire}},
	publisher = {Technische Universiteit Eindhoven},
	author = {IJsselsteijn, W.A. and de Kort, Y.A.W. and Poels, K.},
	year = {2013},
}

@article{schubert_experience_2001,
	title = {The {Experience} of {Presence}: {Factor} {Analytic} {Insights}},
	volume = {10},
	shorttitle = {The {Experience} of {Presence}},
	url = {https://doi.org/10.1162/105474601300343603},
	doi = {10.1162/105474601300343603},
	number = {3},
	urldate = {2024-04-19},
	journal = {Presence: Teleoperators and Virtual Environments},
	author = {Schubert, Thomas and Friedmann, Frank and Regenbrecht, Holger},
	month = jun,
	year = {2001},
	pages = {266--281},
}

@article{usoh_using_2000,
	title = {Using {Presence} {Questionnaires} in {Reality}},
	volume = {9},
	issn = {1054-7460},
	url = {https://ieeexplore.ieee.org/abstract/document/6787863},
	doi = {10.1162/105474600566989},
	number = {5},
	urldate = {2024-04-19},
	journal = {Presence},
	author = {Usoh, Martin and Catena, Ernest and Arman, Sima and Slater, Mel},
	month = oct,
	year = {2000},
	pages = {497--503},
}

@article{witmer_measuring_1998,
	title = {Measuring {Presence} in {Virtual} {Environments}: {A} {Presence} {Questionnaire}},
	volume = {7},
	shorttitle = {Measuring {Presence} in {Virtual} {Environments}},
	url = {https://doi.org/10.1162/105474698565686},
	doi = {10.1162/105474698565686},
	number = {3},
	urldate = {2024-04-19},
	journal = {Presence: Teleoperators and Virtual Environments},
	author = {Witmer, Bob G. and Singer, Michael J.},
	month = jun,
	year = {1998},
	pages = {225--240},
}

@inbook{hart_development_1988,
	series = {Human {Mental} {Workload}},
	title = {Development of {NASA}-{TLX} ({Task} {Load} {Index}): {Results} of {Empirical} and {Theoretical} {Research}},
	volume = {52},
	shorttitle = {Development of {NASA}-{TLX} ({Task} {Load} {Index})},
	url = {https://www.sciencedirect.com/science/article/pii/S0166411508623869},
	urldate = {2024-04-19},
	booktitle = {Advances in {Psychology}},
	publisher = {North-Holland},
	author = {Hart, Sandra G. and Staveland, Lowell E.},
	editor = {Hancock, Peter A. and Meshkati, Najmedin},
	month = jan,
	year = {1988},
	doi = {10.1016/S0166-4115(08)62386-9},
	pages = {139--183},
}

@article{busselle_measuring_2009,
	title = {Measuring {Narrative} {Engagement}},
	volume = {12},
	issn = {1521-3269},
	url = {https://doi.org/10.1080/15213260903287259},
	doi = {10.1080/15213260903287259},
	number = {4},
	urldate = {2024-04-19},
	journal = {Media Psychology},
	author = {Busselle, Rick and Bilandzic, Helena},
	month = nov,
	year = {2009},
	note = {Publisher: Routledge
\_eprint: https://doi.org/10.1080/15213260903287259},
	pages = {321--347},
}

@INPROCEEDINGS{mal2022virtual,
  author={Mal, David and Wolf, Erik and Döllinger, Nina and Botsch, Mario and Wienrich, Carolin and Latoschik, Marc Erich},
  booktitle={2022 IEEE Conference on Virtual Reality and 3D User Interfaces Abstracts and Workshops (VRW)}, 
  title={Virtual Human Coherence and Plausibility – Towards a Validated Scale}, 
  year={2022},
  volume={},
  number={},
  pages={788-789},
  doi={10.1109/VRW55335.2022.00245}}

@article{kim_virtual_2018,
	title = {Virtual reality sickness questionnaire ({VRSQ}): {Motion} sickness measurement index in a virtual reality environment},
	volume = {69},
	issn = {0003-6870},
	shorttitle = {Virtual reality sickness questionnaire ({VRSQ})},
	url = {https://www.sciencedirect.com/science/article/pii/S000368701730282X},
	doi = {10.1016/j.apergo.2017.12.016},
	urldate = {2024-04-19},
	journal = {Applied Ergonomics},
	author = {Kim, Hyun K. and Park, Jaehyun and Choi, Yeongcheol and Choe, Mungyeong},
	month = may,
	year = {2018},
	pages = {66--73},
}

@article{norman2013geq,
  title={Geq (game engagement/experience questionnaire): a review of two papers},
  author={Norman, Kent L},
  journal={Interacting with computers},
  volume={25},
  number={4},
  pages={278--283},
  year={2013},
  publisher={Oxford University Press}
}

@article{kennedy_simulator_1993,
	title = {Simulator {Sickness} {Questionnaire}: {An} {Enhanced} {Method} for {Quantifying} {Simulator} {Sickness}},
	volume = {3},
	issn = {1050-8414},
	shorttitle = {Simulator {Sickness} {Questionnaire}},
	url = {https://doi.org/10.1207/s15327108ijap0303_3},
	doi = {10.1207/s15327108ijap0303_3},
	number = {3},
	urldate = {2024-04-19},
	journal = {The International Journal of Aviation Psychology},
	author = {Kennedy, Robert S. and Lane, Norman E. and Berbaum, Kevin S. and Lilienthal, Michael G.},
	month = jul,
	year = {1993},
	note = {Publisher: Taylor \& Francis
\_eprint: https://doi.org/10.1207/s15327108ijap0303\_3},
	pages = {203--220},
}

@book{vorderer_mec_2004,
	title = {{MEC} spatial presence questionnaire ({MEC}-{SPQ}, {English} and {German} version): {Short} documentation and instructions for application},
	shorttitle = {{MEC} spatial presence questionnaire ({MEC}-{SPQ}, {English} and {German} version)},
	author = {Vorderer, Peter and Wirth, Werner and Gouveia, Feliz and Biocca, Frank and Saari, Timo and Jäncke, Lutz and Böcking, Saskia and Schramm, Holger and Gysbers, Andre and Hartmann, Tilo and Klimmt, Christoph and Laarni, Jari and Ravaja, Niklas and Sacau, Ana and Baumgartner, Thomas and Jäncke, Petra},
	month = jun,
	year = {2004},
	doi = {10.13140/RG.2.2.26232.42249},
	note = {Journal Abbreviation: Report to the European Community, Project Presence: MEC (IST-2001-37661)
Publication Title: Report to the European Community, Project Presence: MEC (IST-2001-37661)},
}

@article{bradley_measuring_1994,
	title = {Measuring emotion: {The} self-assessment manikin and the semantic differential},
	volume = {25},
	issn = {0005-7916},
	shorttitle = {Measuring emotion},
	url = {https://www.sciencedirect.com/science/article/pii/0005791694900639},
	doi = {10.1016/0005-7916(94)90063-9},
	number = {1},
	urldate = {2024-04-20},
	journal = {Journal of Behavior Therapy and Experimental Psychiatry},
	author = {Bradley, Margaret M. and Lang, Peter J.},
	month = mar,
	year = {1994},
	pages = {49--59},
}

@article{thompson_development_2007,
	title = {Development and {Validation} of an {Internationally} {Reliable} {Short}-{Form} of the {Positive} and {Negative} {Affect} {Schedule} ({PANAS})},
	volume = {38},
	issn = {0022-0221},
	url = {https://doi.org/10.1177/0022022106297301},
	doi = {10.1177/0022022106297301},
	language = {en},
	number = {2},
	urldate = {2024-04-20},
	journal = {Journal of Cross-Cultural Psychology},
	author = {Thompson, Edmund R.},
	month = mar,
	year = {2007},
	note = {Publisher: SAGE Publications Inc},
	pages = {227--242},
}

@article{ryan_motivational_2006,
	title = {The {Motivational} {Pull} of {Video} {Games}: {A} {Self}-{Determination} {Theory} {Approach}},
	volume = {30},
	issn = {1573-6644},
	shorttitle = {The {Motivational} {Pull} of {Video} {Games}},
	url = {https://doi.org/10.1007/s11031-006-9051-8},
	doi = {10.1007/s11031-006-9051-8},
	language = {en},
	number = {4},
	urldate = {2024-04-20},
	journal = {Motivation and Emotion},
	author = {Ryan, Richard M. and Rigby, C. Scott and Przybylski, Andrew},
	month = dec,
	year = {2006},
	pages = {344--360},
}

@article{ryan_self-determination_2000,
	title = {Self-determination theory and the facilitation of intrinsic motivation, social development, and well-being},
	volume = {55},
	issn = {1935-990X},
	doi = {10.1037/0003-066X.55.1.68},
	number = {1},
	journal = {American Psychologist},
	author = {Ryan, Richard M. and Deci, Edward L.},
	year = {2000},
	note = {Place: US
Publisher: American Psychological Association},
	pages = {68--78},
}

@inproceedings{laugwitz_construction_2008,
	address = {Berlin, Heidelberg},
	title = {Construction and {Evaluation} of a {User} {Experience} {Questionnaire}},
	isbn = {978-3-540-89350-9},
	doi = {10.1007/978-3-540-89350-9_6},
	language = {en},
	booktitle = {{HCI} and {Usability} for {Education} and {Work}},
	publisher = {Springer},
	author = {Laugwitz, Bettina and Held, Theo and Schrepp, Martin},
	editor = {Holzinger, Andreas},
	year = {2008},
	pages = {63--76},
}

@inproceedings{lin_effects_2002,
	title = {Effects of field of view on presence, enjoyment, memory, and simulator sickness in a virtual environment},
	url = {https://ieeexplore.ieee.org/abstract/document/996519},
	doi = {10.1109/VR.2002.996519},
	urldate = {2024-04-20},
	booktitle = {Proceedings {IEEE} {Virtual} {Reality} 2002},
	author = {Lin, J.J.-W. and Duh, H.B.L. and Parker, D.E. and Abi-Rached, H. and Furness, T.A.},
	month = mar,
	year = {2002},
	note = {ISSN: 1087-8270},
	pages = {164--171},
}

@article{jennett_measuring_2008,
	title = {Measuring and defining the experience of immersion in games},
	volume = {66},
	issn = {1071-5819},
	url = {https://www.sciencedirect.com/science/article/pii/S1071581908000499},
	doi = {10.1016/j.ijhcs.2008.04.004},
	number = {9},
	urldate = {2024-04-20},
	journal = {International Journal of Human-Computer Studies},
	author = {Jennett, Charlene and Cox, Anna L. and Cairns, Paul and Dhoparee, Samira and Epps, Andrew and Tijs, Tim and Walton, Alison},
	month = sep,
	year = {2008},
	pages = {641--661},
}

@article{gonzalez-franco_avatar_2018,
	title = {Avatar {Embodiment}. {Towards} a {Standardized} {Questionnaire}},
	volume = {5},
	issn = {2296-9144},
	url = {https://www.frontiersin.org/articles/10.3389/frobt.2018.00074},
	doi = {10.3389/frobt.2018.00074},
	language = {English},
	urldate = {2024-04-20},
	journal = {Frontiers in Robotics and AI},
	author = {Gonzalez-Franco, Mar and Peck, Tabitha C.},
	month = jun,
	year = {2018},
	note = {Publisher: Frontiers},
}

@incollection{brooke_sus_1996,
	title = {{SUS}: {A} '{Quick} and {Dirty}' {Usability} {Scale}},
	isbn = {978-0-429-15701-1},
	shorttitle = {{SUS}},
	booktitle = {Usability {Evaluation} {In} {Industry}},
	publisher = {CRC Press},
	author = {Brooke, John},
	year = {1996},
	note = {Num Pages: 6},
}

@article{gachter_measuring_2015,
	title = {Measuring the {Closeness} of {Relationships}: {A} {Comprehensive} {Evaluation} of the '{Inclusion} of the {Other} in the {Self}' {Scale}},
	volume = {10},
	issn = {1932-6203},
	shorttitle = {Measuring the {Closeness} of {Relationships}},
	url = {https://journals.plos.org/plosone/article?id=10.1371/journal.pone.0129478},
	doi = {10.1371/journal.pone.0129478},
	language = {en},
	number = {6},
	urldate = {2024-04-20},
	journal = {PLOS ONE},
	author = {Gächter, Simon and Starmer, Chris and Tufano, Fabio},
	month = jun,
	year = {2015},
	note = {Publisher: Public Library of Science},
	pages = {e0129478},
}

@inproceedings{zhu2023can,
  title={“Can You Move It?”: The Design and Evaluation of Moving VR Shots in Sport Broadcast},
  author={Zhu, Xiuqi and Wang, Chenyi and Guo, Zichun and Zhao, Yifan and Jiao, Yang},
  booktitle={2023 IEEE International Symposium on Mixed and Augmented Reality (ISMAR)},
  pages={839--848},
  year={2023},
  organization={IEEE}
}

@inproceedings{xu2023user,
  title={User Experience of Collaborative Co-located Mixed Reality: a User Study in Teaching Veterinary Radiation Safety Rules},
  author={Xu, Xuanhui and Puggioni, Antonella and Kilroy, David and Campbell, Abraham G},
  booktitle={2023 IEEE International Symposium on Mixed and Augmented Reality (ISMAR)},
  pages={583--590},
  year={2023},
  organization={IEEE}
}

@inproceedings{lee2018user,
  title={A user study on mr remote collaboration using live 360 video},
  author={Lee, Gun A and Teo, Theophilus and Kim, Seungwon and Billinghurst, Mark},
  booktitle={2018 IEEE International Symposium on Mixed and Augmented Reality (ISMAR)},
  pages={153--164},
  year={2018},
  organization={IEEE}
}

@inproceedings{thanyadit2019observar,
  title={ObserVAR: Visualization system for observing virtual reality users using augmented reality},
  author={Thanyadit, Santawat and Punpongsanon, Parinya and Pong, Ting-Chuen},
  booktitle={2019 IEEE International Symposium on Mixed and Augmented Reality (ISMAR)},
  pages={258--268},
  year={2019},
  organization={IEEE}
}

@inproceedings{gattullo2022biophilic,
  title={Biophilic enriched virtual environments for industrial training: a user study},
  author={Gattullo, Michele and Laviola, Enricoandrea and Romano, Sara and Evangelista, Alessandro and Manghisi, Vito Modesto and Fiorentino, Michele and Uva, Antonio Emmanuele},
  booktitle={2022 IEEE international symposium on mixed and augmented reality (ISMAR)},
  pages={206--214},
  year={2022},
  organization={IEEE}
}

@INPROCEEDINGS{Ito_2023_wind,
  author={Ito, Kenichi and Hosoi, Juro and Ban, Yuki and Kikuchi, Takayuki and Nakagawa, Kyosuke and Kitagawa, Hanako and Murakami, Chizuru and Imai, Yosuke and Warisawa, Shin'ichi},
  booktitle={2023 IEEE Conference Virtual Reality and 3D User Interfaces (VR)}, 
  title={Wind comfort and emotion can be changed by the cross-modal presentation of audio-visual stimuli of indoor and outdoor environments}, 
  year={2023},
  volume={},
  number={},
  pages={215-225},
  doi={10.1109/VR55154.2023.00037}}

@book{zijlstra1985construction,
  title={The construction of a scale to measure perceived effort. Report},
  author={Zijlstra, FRH and Van Doorn, L},
  year={1985},
  publisher={Delft University of Technology, Delft}
}

@inproceedings{kwon2022infinite,
  title={Infinite virtual space exploration using space tiling and perceivable reset at fixed positions},
  author={Kwon, Soon-Uk and Jeon, Sang-Bin and Hwang, June-Young and Cho, Yong-Hun and Park, Jinhyung and Lee, In-Kwon},
  booktitle={2022 IEEE International Symposium on Mixed and Augmented Reality (ISMAR)},
  pages={758--767},
  year={2022},
  organization={IEEE}
}

@inproceedings{venkatakrishnan2020comparative,
  title={Comparative evaluation of the effects of motion control on cybersickness in immersive virtual environments},
  author={Venkatakrishnan, Roshan and Venkatakrishnan, Rohith and Bhargava, Ayush and Lucaites, Kathryn and Solini, Hannah and Volonte, Matias and Robb, Andrew and Babu, Sabarish V and Lin, Wen-Chieh and Lin, Yun-Xuan},
  booktitle={2020 IEEE Conference on Virtual Reality and 3D User Interfaces (VR)},
  pages={672--681},
  year={2020},
  organization={IEEE}
}

@INPROCEEDINGS{Venkatakrishnan_2020_Equation,
  author={Venkatakrishnan, Rohith and Venkatakrishnan, Roshan and Anaraky, Reza Ghaiumy and Volonte, Matias and Knijnenburg, Bart and Babu, Sabarish V},
  booktitle={2020 IEEE Conference on Virtual Reality and 3D User Interfaces (VR)}, 
  title={A Structural Equation Modeling Approach to Understand the Relationship between Control, Cybersickness and Presence in Virtual Reality}, 
  year={2020},
  volume={},
  number={},
  pages={682-691},
  doi={10.1109/VR46266.2020.00091}}

@inproceedings{pointecker2022bridging,
  title={Bridging the gap across realities: Visual transitions between virtual and augmented reality},
  author={Pointecker, Fabian and Friedl, Judith and Schwajda, Daniel and Jetter, Hans-Christian and Anthes, Christoph},
  booktitle={2022 IEEE international symposium on mixed and augmented reality (ISMAR)},
  pages={827--836},
  year={2022},
  organization={IEEE}
}

@inproceedings{giovannelli2022exploring,
  title={Exploring the impact of visual information on intermittent typing in virtual reality},
  author={Giovannelli, Alexander and Lisle, Lee and Bowman, Doug A},
  booktitle={2022 IEEE International Symposium on Mixed and Augmented Reality (ISMAR)},
  pages={8--17},
  year={2022},
  organization={IEEE}
}

@article{schrepp2017design,
  title={Design and evaluation of a short version of the user experience questionnaire (UEQ-S)},
  author={Schrepp, Martin and Hinderks, Andreas and Thomaschewski, J{\"o}rg},
  journal={International Journal of Interactive Multimedia and Artificial Intelligence, 4 (6), 103-108.},
  year={2017},
  publisher={UNIR}
}

@ARTICLE{Chittaro2015assessing,
  author={Chittaro, Luca and Buttussi, Fabio},
  journal={IEEE Transactions on Visualization and Computer Graphics}, 
  title={Assessing Knowledge Retention of an Immersive Serious Game vs. a Traditional Education Method in Aviation Safety}, 
  year={2015},
  volume={21},
  number={4},
  pages={529-538},
  doi={10.1109/TVCG.2015.2391853}}

@inproceedings{shin2021user,
  title={A user-oriented approach to space-adaptive augmentation: The effects of spatial affordance on narrative experience in an augmented reality detective game},
  author={Shin, Jae-eun and Yoon, Boram and Kim, Dooyoung and Woo, Woontack},
  booktitle={Proceedings of the 2021 CHI Conference on Human Factors in Computing Systems},
  pages={1--13},
  year={2021}
}

@inproceedings{liu2022investigating,
  title={Investigating the effects of leading and following behaviors of virtual humans in collaborative fine motor tasks in virtual reality},
  author={Liu, Kuan-Yu and Wong, Sai-Keung and Volonte, Matias and Ebrahimi, Elham and Babu, Sabarish V},
  booktitle={2022 IEEE Conference on Virtual Reality and 3D User Interfaces (VR)},
  pages={330--339},
  year={2022},
  organization={IEEE}
}

@inproceedings{thoravi2022dreamstream,
  title={Dreamstream: Immersive and interactive spectating in vr},
  author={Thoravi Kumaravel, Balasaravanan and Wilson, Andrew D},
  booktitle={Proceedings of the 2022 CHI Conference on Human Factors in Computing Systems},
  pages={1--17},
  year={2022}
}

@inproceedings{chalil2011synchronous,
  title={Synchronous remote usability testing: a new approach facilitated by virtual worlds},
  author={Chalil Madathil, Kapil and Greenstein, Joel S},
  booktitle={Proceedings of the SIGCHI Conference on Human Factors in Computing Systems},
  pages={2225--2234},
  year={2011}
}

@inproceedings{li2020analysing,
  title={Analysing usability and presence of a virtual reality operating room (VOR) simulator during laparoscopic surgery training},
  author={Li, Meng and Ganni, Sandeep and Ponten, Jeroen and Albayrak, Armagan and Rutkowski, Anne-F and Jakimowicz, Jack},
  booktitle={2020 IEEE Conference on Virtual Reality and 3D User Interfaces (VR)},
  pages={566--572},
  year={2020},
  organization={IEEE}
}

@inproceedings{ricca2020influence,
  title={Influence of hand visualization on tool-based motor skills training in an immersive VR simulator},
  author={Ricca, Aylen and Chellali, Amine and Otmane, Samir},
  booktitle={2020 IEEE international symposium on mixed and augmented reality (ISMAR)},
  pages={260--268},
  year={2020},
  organization={IEEE}
}

@inproceedings{cai2020thermairglove,
  title={Thermairglove: A pneumatic glove for thermal perception and material identification in virtual reality},
  author={Cai, Shaoyu and Ke, Pingchuan and Narumi, Takuji and Zhu, Kening},
  booktitle={2020 IEEE conference on virtual reality and 3D user interfaces (VR)},
  pages={248--257},
  year={2020},
  organization={IEEE}
}

@article{skarbez2020immersion,
  title={Immersion and coherence: Research agenda and early results},
  author={Skarbez, Richard and Brooks, Frederick P and Whitton, Mary C},
  journal={IEEE transactions on visualization and computer graphics},
  volume={27},
  number={10},
  pages={3839--3850},
  year={2020},
  publisher={IEEE}
}

@inproceedings{gandy2010experiences,
  title={Experiences with an AR evaluation test bed: Presence, performance, and physiological measurement},
  author={Gandy, Maribeth and Catrambone, Richard and MacIntyre, Blair and Alvarez, Chris and Eiriksdottir, Elsa and Hilimire, Matthew and Davidson, Brian and McLaughlin, Anne Collins},
  booktitle={2010 IEEE International Symposium on Mixed and Augmented Reality},
  pages={127--136},
  year={2010},
  organization={IEEE}
}

@inproceedings{grubert2018text,
  title={Text entry in immersive head-mounted display-based virtual reality using standard keyboards},
  author={Grubert, Jens and Witzani, Lukas and Ofek, Eyal and Pahud, Michel and Kranz, Matthias and Kristensson, Per Ola},
  booktitle={2018 IEEE Conference on Virtual Reality and 3D User Interfaces (VR)},
  pages={159--166},
  year={2018},
  organization={IEEE}
}

@inproceedings{boldt2018you,
  title={You shall not pass: Non-intrusive feedback for virtual walls in VR environments with room-scale mapping},
  author={Boldt, Mette and Bonfert, Michael and Lehne, Inga and Cahnbley, Melina and Korschinq, Kim and Bikas, Loannis and Finke, Stefan and Hanci, Martin and Kraft, Valentin and Liu, Boxuan and others},
  booktitle={2018 IEEE Conference on Virtual Reality and 3D User Interfaces (VR)},
  pages={143--150},
  year={2018},
  organization={IEEE}
}

@inproceedings{zhao2023leanon,
  title={LeanOn: Simulating Balance Vehicle Locomotion in Virtual Reality},
  author={Zhao, Ziyue and Li, Yue and Liang, Hai-Ning},
  booktitle={2023 IEEE International Symposium on Mixed and Augmented Reality (ISMAR)},
  pages={415--424},
  year={2023},
  organization={IEEE}
}

@inproceedings{kim2023active,
  title={Active Engagement with Virtual Reality Reduces Stress and Increases Positive Emotions},
  author={Kim, Irene and Azimi, Ehsan and Kazanzides, Peter and Huang, Chien-Ming},
  booktitle={2023 IEEE International Symposium on Mixed and Augmented Reality (ISMAR)},
  pages={523--532},
  year={2023},
  organization={IEEE}
}

@inproceedings{krum2018influences,
  title={Influences on the elicitation of interpersonal space with virtual humans},
  author={Krum, David M and Kang, Sin-Hwa and Phan, Thai},
  booktitle={2018 IEEE Conference on Virtual Reality and 3D User Interfaces (VR)},
  pages={223--9},
  year={2018},
  organization={IEEE}
}

@INPROCEEDINGS{chen2024effects,
  author={Chen, Yi-An and Wong, Sai-Keung and Chao, Yu-Ting and Babu, Sabarish V.},
  booktitle={2024 IEEE International Symposium on Mixed and Augmented Reality (ISMAR)}, 
  title={Effects of Organizational and Behavioral Reactions of Virtual Crowds on Users’ Affect and Behavior in a Simulated Stressful Situation}, 
  year={2024},
  volume={},
  number={},
  pages={1107-1116},
  doi={10.1109/ISMAR62088.2024.00127}}

@INPROCEEDINGS{jing2024superpowering,
  author={Jing, Allison and Teo, Theophilus and McDade, Jeremy and Zhang, Chenkai and Wang, Yi and Mitrofan, Andrei and Thareja, Rushil and Shin, Heesook and Lee, Yongho and Gil, Youn-Hee and Billinghurst, Mark and Lee, Gun A.},
  booktitle={2024 IEEE International Symposium on Mixed and Augmented Reality (ISMAR)}, 
  title={Superpowering Emotion Through Multimodal Cues in Collaborative VR}, 
  year={2024},
  volume={},
  number={},
  pages={160-169},
  doi={10.1109/ISMAR62088.2024.00030}}

@ARTICLE{xiong2024petpresence,
  author={Xiong, Ningchang and Liu, Qingqin and Zhu, Kening},
  journal={IEEE Transactions on Visualization and Computer Graphics}, 
  title={PetPresence: Investigating the Integration of Real-World Pet Activities in Virtual Reality}, 
  year={2024},
  volume={30},
  number={5},
  pages={2559-2569},
  doi={10.1109/TVCG.2024.3372095}}

@inproceedings{Bonnail2024real,
author = {Bonnail, Elise and Frommel, Julian and Lecolinet, Eric and Huron, Samuel and Gugenheimer, Jan},
title = {Was it Real or Virtual? Confirming the Occurrence and Explaining Causes of Memory Source Confusion between Reality and Virtual Reality},
year = {2024},
isbn = {9798400703300},
publisher = {Association for Computing Machinery},
address = {New York, NY, USA},
url = {https://doi.org/10.1145/3613904.3641992},
doi = {10.1145/3613904.3641992},
booktitle = {Proceedings of the 2024 CHI Conference on Human Factors in Computing Systems},
articleno = {796},
numpages = {17},
location = {Honolulu, HI, USA},
series = {CHI '24}
}

@inproceedings{raeburn2021varying,
  title={Varying user agency and interaction opportunities in a home mobile augmented virtuality story},
  author={Raeburn, Gideon and Tokarchuk, Laurissa},
  booktitle={2021 IEEE International Symposium on Mixed and Augmented Reality (ISMAR)},
  pages={347--356},
  year={2021},
  organization={IEEE}
}

@ARTICLE{shin2023effects,
  author={Shin, Jae-eun and Yoon, Boram and Kim, Dooyoung and Woo, Woontack},
  journal={IEEE Transactions on Visualization and Computer Graphics}, 
  title={The Effects of Spatial Complexity on Narrative Experience in Space-Adaptive AR Storytelling}, 
  year={2023},
  volume={29},
  number={12},
  pages={5137-5148},
  doi={10.1109/TVCG.2022.3201934}
}

@inproceedings{belani2023investigating,
  title={Investigating spatial representation of learning content in virtual reality learning environments},
  author={Belani, Manshul and Singh, Harsh Vardhan and Parnami, Aman and Singh, Pushpendra},
  booktitle={2023 IEEE Conference Virtual Reality and 3D User Interfaces (VR)},
  pages={33--43},
  year={2023},
  organization={IEEE}
}

@inproceedings{schott2023vreal,
  title={Is this the vReal Life? Manipulating Visual Fidelity of Immersive Environments for Medical Task Simulation},
  author={Schott, Danny and Heinrich, Florian and Stallmeister, Lara and Moritz, Julia and Hensen, Bennet and Hansen, Christian},
  booktitle={2023 IEEE International Symposium on Mixed and Augmented Reality (ISMAR)},
  pages={1171--1180},
  year={2023},
  organization={IEEE}
}

@article{schmitz2018you,
  title={You spin my head right round: Threshold of limited immersion for rotation gains in redirected walking},
  author={Schmitz, Patric and Hildebrandt, Julian and Valdez, Andr{\'e} Calero and Kobbelt, Leif and Ziefle, Martina},
  journal={IEEE transactions on visualization and computer graphics},
  volume={24},
  number={4},
  pages={1623--1632},
  year={2018},
  publisher={IEEE}
}

@article{Buck2022TheIO,
  title={The Impact of Embodiment and Avatar Sizing on Personal Space in Immersive Virtual Environments},
  author={Lauren E. Buck and Soumyajit Chakraborty and Bobby Bodenheimer},
  journal={IEEE Transactions on Visualization and Computer Graphics},
  year={2022},
  volume={28},
  pages={2102-2113},
  url={https://api.semanticscholar.org/CorpusID:246866577}
}

@article{ganias2023comparing,
  title={Comparing Different Grasping Visualizations for Object Manipulation in VR using Controllers},
  author={Ganias, Giorgos and Lougiakis, Christos and Katifori, Akrivi and Roussou, Maria and Ioannidis, Yannis and others},
  journal={IEEE Transactions on Visualization and Computer Graphics},
  volume={29},
  number={5},
  pages={2369--2378},
  year={2023},
  publisher={IEEE}
}

@article{nie2019analysis,
  title={Analysis on mitigation of visually induced motion sickness by applying dynamical blurring on a user's retina},
  author={Nie, Guang-Yu and Duh, Henry Been-Lirn and Liu, Yue and Wang, Yongtian},
  journal={IEEE transactions on visualization and computer graphics},
  volume={26},
  number={8},
  pages={2535--2545},
  year={2019},
  publisher={IEEE}
}

@inproceedings{lugrin2018any,
  title={Any “body” there? avatar visibility effects in a virtual reality game},
  author={Lugrin, Jean-Luc and Ertl, Maximilian and Krop, Philipp and Kl{\"u}pfel, Richard and Stierstorfer, Sebastian and Weisz, Bianka and R{\"u}ck, Maximilian and Schmitt, Johann and Schmidt, Nina and Latoschik, Marc Erich},
  booktitle={2018 IEEE conference on virtual reality and 3D user interfaces (VR)},
  pages={17--24},
  year={2018},
  organization={IEEE}
}

@inproceedings{monteiro2020depth,
  title={An in-depth exploration of the effect of 2d/3d views and controller types on first person shooter games in virtual reality},
  author={Monteiro, Diego and Liang, Hai-Ning and Wang, Jialin and Chen, Hao and Baghaei, Nilufar},
  booktitle={2020 IEEE International Symposium on Mixed and Augmented Reality (ISMAR)},
  pages={713--724},
  year={2020},
  organization={IEEE}
}

@article{kim2023or,
  title={``To Be or Not to Be Me?'': Exploration of Self-Similar Effects ofAvatars on Social Virtual Reality Experiences},
  author={Kim, Hayeon and Park, Jinhyung and Lee, In-Kwon},
  journal={IEEE Transactions on Visualization and Computer Graphics},
  year={2023},
  publisher={IEEE}
}

@inproceedings{bhargava2023empirically,
  title={Empirically evaluating the effects of eye height and self-avatars on dynamic passability affordances in virtual reality},
  author={Bhargava, Ayush and Venkatakrishnan, Roshan and Venkatakrishnan, Rohith and Solini, Hannah and Lucaites, Kathryn and Robb, Andrew C and Pagano, Christopher C and Babu, Sabarish V},
  booktitle={2023 IEEE Conference Virtual Reality and 3D User Interfaces (VR)},
  pages={308--317},
  year={2023},
  organization={IEEE}
}

@article{venkatakrishnan2023virtual,
  title={How virtual hand representations affect the perceptions of dynamic affordances in virtual reality},
  author={Venkatakrishnan, Roshan and Venkatakrishnan, Rohith and Raveendranath, Balagopal and Pagano, Christopher C and Robb, Andrew C and Lin, Wen-Chieh and Babu, Sabarish V},
  journal={IEEE Transactions on Visualization and Computer Graphics},
  volume={29},
  number={5},
  pages={2258--2268},
  year={2023},
  publisher={IEEE}
}

@inproceedings{whitlock2018interacting,
  title={Interacting with distant objects in augmented reality},
  author={Whitlock, Matt and Harnner, Ethan and Brubaker, Jed R and Kane, Shaun and Szafir, Danielle Albers},
  booktitle={2018 IEEE Conference on Virtual Reality and 3D User Interfaces (VR)},
  pages={41--48},
  year={2018},
  organization={IEEE}
}

@inproceedings{liao2020data,
  title={Data-driven spatio-temporal analysis via multi-modal zeitgebers and cognitive load in VR},
  author={Liao, Haodong and Xie, Ning and Li, Huiyuan and Li, Yuhang and Su, Jianping and Jiang, Feng and Huang, Weipeng and Shen, Heng Tao},
  booktitle={2020 IEEE Conference on Virtual Reality and 3D User Interfaces (VR)},
  pages={473--482},
  year={2020},
  organization={IEEE}
}

@inproceedings{mahmood2019improving,
  title={Improving information sharing and collaborative analysis for remote geospatial visualization using mixed reality},
  author={Mahmood, Tahir and Fulmer, Willis and Mungoli, Neelesh and Huang, Jian and Lu, Aidong},
  booktitle={2019 IEEE International Symposium on Mixed and Augmented Reality (ISMAR)},
  pages={236--247},
  year={2019},
  organization={IEEE}
}

@ARTICLE{cauquis2024investigating,
  author={Cauquis, Julien and Peillard, Etienne and Dominjon, Lionel and Duval, Thierry and Moreau, Guillaume},
  journal={IEEE Transactions on Visualization and Computer Graphics}, 
  title={Investigating Whether the Mass of a Tool Replica Influences Virtual Training Learning Outcomes}, 
  year={2024},
  volume={30},
  number={5},
  pages={2411-2421},
  doi={10.1109/TVCG.2024.3372041}}

@INPROCEEDINGS{ahmmed2024system,
  author={Ahmmed, Asif and Butts, Erica and Naeiji, Kimia and Thiamwong, Ladda and Daher, Salam},
  booktitle={2024 IEEE International Symposium on Mixed and Augmented Reality (ISMAR)}, 
  title={System Usability and Technology Acceptance of a Geriatric Embodied Virtual Human Simulation in Augmented Reality}, 
  year={2024},
  volume={},
  number={},
  pages={594-603},
  doi={10.1109/ISMAR62088.2024.00074}}

@inproceedings{amaro2022design,
  title={Design and evaluation of travel and orientation techniques for desk vr},
  author={Amaro, Guilherme and Mendes, Daniel and Rodrigues, Rui},
  booktitle={2022 IEEE Conference on Virtual Reality and 3D User Interfaces (VR)},
  pages={222--231},
  year={2022},
  organization={IEEE}
}

@article{xu2023gesturesurface,
  title={GestureSurface: VR Sketching through Assembling Scaffold Surface with Non-Dominant Hand},
  author={Xu, Xinchi and Zhou, Yang and Shao, Bingchan and Feng, Guihuan and Yu, Chun},
  journal={IEEE Transactions on Visualization and Computer Graphics},
  volume={29},
  number={5},
  pages={2499--2507},
  year={2023},
  publisher={IEEE}
}

@inproceedings{kanamori2018obstacle,
  title={Obstacle avoidance method in real space for virtual reality immersion},
  author={Kanamori, Kohei and Sakata, Nobuchika and Tominaga, Tomu and Hijikata, Yoshinori and Harada, Kensuke and Kiyokawa, Kiyoshi},
  booktitle={2018 IEEE International Symposium on Mixed and Augmented Reality (ISMAR)},
  pages={80--89},
  year={2018},
  organization={IEEE}
}

@inproceedings{mathis2022virtual,
  title={Virtual reality observations: Using virtual reality to augment lab-based shoulder surfing research},
  author={Mathis, Florian and O’Hagan, Joseph and Khamis, Mohamed and Vaniea, Kami},
  booktitle={2022 IEEE Conference on Virtual Reality and 3D User Interfaces (VR)},
  pages={291--300},
  year={2022},
  organization={IEEE}
}

@INPROCEEDINGS{kim2024audience,
  author={Kim, You-Jin and Sra, Misha and Höllerer, Tobias},
  booktitle={2024 IEEE International Symposium on Mixed and Augmented Reality (ISMAR)}, 
  title={Audience Amplified: Virtual Audiences in Asynchronously Performed AR Theater}, 
  year={2024},
  volume={},
  number={},
  pages={475-484},
  doi={10.1109/ISMAR62088.2024.00062}}

@inproceedings{fan2024spinshot,
author = {Fan, Chia-An and Wu, En-Huei and Cheng, Chia-Yu and Chang, Yu-Cheng and Lopez, Alvaro and Chen, Yu and Chi, Chia-Chen and Chan, Yi-Sheng and Tsai, Ching-Yi and Chen, Mike Y.},
title = {SpinShot: Optimizing Both Physical and Perceived Force Feedback of Flywheel-Based, Directional Impact Handheld Devices},
year = {2024},
isbn = {9798400706288},
publisher = {Association for Computing Machinery},
address = {New York, NY, USA},
url = {https://doi.org/10.1145/3654777.3676433},
doi = {10.1145/3654777.3676433},
booktitle = {Proceedings of the 37th Annual ACM Symposium on User Interface Software and Technology},
articleno = {138},
numpages = {15},
location = {Pittsburgh, PA, USA},
series = {UIST '24}
}

@inproceedings{kim2024QuadStretcher,
author = {Kim, Taejun and Shim, Youngbo Aram and Kim, Youngin and Kim, Sunbum and Lee, Jaeyeon and Lee, Geehyuk},
title = {QuadStretcher: A Forearm-Worn Skin Stretch Display for Bare-Hand Interaction in AR/VR},
year = {2024},
isbn = {9798400703300},
publisher = {Association for Computing Machinery},
address = {New York, NY, USA},
url = {https://doi.org/10.1145/3613904.3642067},
doi = {10.1145/3613904.3642067},
booktitle = {Proceedings of the 2024 CHI Conference on Human Factors in Computing Systems},
articleno = {409},
numpages = {15},
location = {Honolulu, HI, USA},
series = {CHI '24}
}

@inproceedings{suma2011leveraging,
  title={Leveraging change blindness for redirection in virtual environments},
  author={Suma, Evan A and Clark, Seth and Krum, David and Finkelstein, Samantha and Bolas, Mark and Warte, Zachary},
  booktitle={2011 IEEE Virtual Reality Conference},
  pages={159--166},
  year={2011},
  organization={IEEE}
}

@inproceedings{dey2020neurophysiological,
  title={A neurophysiological approach for measuring presence in immersive virtual environments},
  author={Dey, Arindam and Phoon, Jane and Saha, Shuvodeep and Dobbins, Chelsea and Billinghurst, Mark},
  booktitle={2020 IEEE international symposium on mixed and augmented reality (Ismar)},
  pages={474--485},
  year={2020},
  organization={IEEE}
}

@article{martin2023study,
  title={A study of change blindness in immersive environments},
  author={Martin, Daniel and Sun, Xin and Gutierrez, Diego and Masia, Belen},
  journal={IEEE Transactions on Visualization and Computer Graphics},
  volume={29},
  number={5},
  pages={2446--2455},
  year={2023},
  publisher={IEEE}
}

@inproceedings{young2006demand,
  title={Demand characteristics of a questionnaire used to assess motion sickness in a virtual environment},
  author={Young, Sean D and Adelstein, Bernard D and Ellis, Stephen R},
  booktitle={IEEE virtual reality conference (VR 2006)},
  pages={97--102},
  year={2006},
  organization={IEEE}
}

@article{young2007demand,
  title={Demand characteristics in assessing motion sickness in a virtual environment: Or does taking a motion sickness questionnaire make you sick?},
  author={Young, Sean D and Adelstein, Bernard D and Ellis, Stephen R},
  journal={IEEE transactions on visualization and computer graphics},
  volume={13},
  number={3},
  pages={422--428},
  year={2007},
  publisher={IEEE}
}

@article{bolte2015subliminal,
  title={Subliminal reorientation and repositioning in immersive virtual environments using saccadic suppression},
  author={Bolte, Benjamin and Lappe, Markus},
  journal={IEEE transactions on visualization and computer graphics},
  volume={21},
  number={4},
  pages={545--552},
  year={2015},
  publisher={IEEE}
}

@ARTICLE{zenner2021combining,
  author={Zenner, André and Ullmann, Kristin and Krüger, Antonio},
  journal={IEEE Transactions on Visualization and Computer Graphics}, 
  title={Combining Dynamic Passive Haptics and Haptic Retargeting for Enhanced Haptic Feedback in Virtual Reality}, 
  year={2021},
  volume={27},
  number={5},
  pages={2627-2637},
  doi={10.1109/TVCG.2021.3067777}}

@inproceedings{mahmud2023auditory,
  title={Auditory, Vibrotactile, or Visual? Investigating the Effective Feedback Modalities to Improve Standing Balance in Immersive Virtual Reality for People with Balance Impairments Due to Type 2 Diabetes},
  author={Mahmud, M Rasel and Cordova, Alberto and Quarles, John},
  booktitle={2023 IEEE International Symposium on Mixed and Augmented Reality (ISMAR)},
  pages={573--582},
  year={2023},
  organization={IEEE}
}

@article{kasahara2016jackin,
  title={Jackin head: Immersive visual telepresence system with omnidirectional wearable camera},
  author={Kasahara, Shunichi and Nagai, Shohei and Rekimoto, Jun},
  journal={IEEE transactions on visualization and computer graphics},
  volume={23},
  number={3},
  pages={1222--1234},
  year={2016},
  publisher={IEEE}
}

@article{mimnaugh2023virtual,
  title={Virtual reality sickness reduces attention during immersive experiences},
  author={Mimnaugh, Katherine J and Center, Evan G and Suomalainen, Markku and Becerra, Israel and Lozano, Eliezer and Murrieta-Cid, Rafael and Ojala, Timo and LaValle, Steven M and Federmeier, Kara D},
  journal={IEEE Transactions on Visualization and Computer Graphics},
  year={2023},
  publisher={IEEE}
}

@inproceedings{kang2023giant,
  title={Giant Finger: A Novel Visuo-Somatosensory Approach to Simulating Lower Body Movements in Virtual Reality},
  author={Kang, Seongjun and Kim, Gwangbin and Kim, SeungJun},
  booktitle={2023 IEEE International Symposium on Mixed and Augmented Reality (ISMAR)},
  pages={233--242},
  year={2023},
  organization={IEEE}
}

@inproceedings{valentini2020improving,
  title={Improving obstacle awareness to enhance interaction in virtual reality},
  author={Valentini, Ivan and Ballestin, Giorgio and Bassano, Chiara and Solari, Fabio and Chessa, Manuela},
  booktitle={2020 IEEE Conference on Virtual Reality and 3D User Interfaces (VR)},
  pages={44--52},
  year={2020},
  organization={IEEE}
}

@inproceedings{heldal2005immersiveness,
  title={Immersiveness and symmetry in copresent scenarios},
  author={Heldal, Ilona and Schroeder, Ralph and Steed, Anthony and Axelsson, A-S and Spant, M and Widestrom, J},
  booktitle={IEEE Proceedings. VR 2005. Virtual Reality, 2005.},
  pages={171--178},
  year={2005},
  organization={IEEE}
}

@inproceedings{chen2023leap,
  title={Leap to the Eye: Implicit Gaze-based Interaction to Reveal Invisible Objects for Virtual Environment Exploration},
  author={Chen, Yang-Sheng and Hsieh, Chiao-En and Jie, Miguel Then Ying and Han, Ping-Hsuan and Hung, Yi-Ping},
  booktitle={2023 IEEE International Symposium on Mixed and Augmented Reality (ISMAR)},
  pages={214--222},
  year={2023},
  organization={IEEE}
}

@INPROCEEDINGS{Kim2024Crossing,
  author={Kim, DongHoon and Han, Dongyun and Bak, Siyeon and Cho, Isaac},
  booktitle={2024 IEEE International Symposium on Mixed and Augmented Reality (ISMAR)}, 
  title={Crossing Rays: Evaluation of Bimanual Mid-air Selection Techniques in an Immersive Environment}, 
  year={2024},
  volume={},
  number={},
  pages={329-338},
  doi={10.1109/ISMAR62088.2024.00047}}

@inproceedings{Lee2024Viewer2Explorer,
author = {Lee, Chaeeun and Kim, Jinwook and Yi, Hyeonbeom and Lee, Woohun},
title = {Viewer2Explorer: Designing a Map Interface for Spatial Navigation in Linear 360 Museum Exhibition Video},
year = {2024},
isbn = {9798400703300},
publisher = {Association for Computing Machinery},
address = {New York, NY, USA},
url = {https://doi.org/10.1145/3613904.3642952},
doi = {10.1145/3613904.3642952},
booktitle = {Proceedings of the 2024 CHI Conference on Human Factors in Computing Systems},
articleno = {400},
numpages = {15},
location = {Honolulu, HI, USA},
series = {CHI '24}
}

@INPROCEEDINGS{wang2024a3rt,
  author={Wang, Xuanyu and Zhang, Weizhan and Fu, Hongbo},
  booktitle={2024 IEEE Conference Virtual Reality and 3D User Interfaces (VR)}, 
  title={A3RT: Attention-Aware AR Teleconferencing with Life-Size 2.5D Video Avatars}, 
  year={2024},
  volume={},
  number={},
  pages={211-221},
  doi={10.1109/VR58804.2024.00044}}

@INPROCEEDINGS{Acevedo2024effects,
  author={Acevedo, Pedro and Choi, Minsoo and Magana, Alejandra J. and Benes, Bedrich and Mousas, Christos},
  booktitle={2024 IEEE International Symposium on Mixed and Augmented Reality (ISMAR)}, 
  title={The Effects of Immersion and Dimensionality in Virtual Reality Science Simulations: The Case of Charged Particles}, 
  year={2024},
  volume={},
  number={},
  pages={170-179},
  doi={10.1109/ISMAR62088.2024.00031}}

@INPROCEEDINGS{zou2024effect,
  author={Zou, Qianyuan and Bai, Huidong and Chang, Zhuang and Xiao, Zirui and Tian, Suizi and Duh, Henry Been-Lirn and Fowler, Allan and Billinghurst, Mark},
  booktitle={2024 IEEE International Symposium on Mixed and Augmented Reality (ISMAR)}, 
  title={The Effect of Interface Types and Immersive Environments on Drawing Accuracy and User Comfort}, 
  year={2024},
  volume={},
  number={},
  pages={836-845},
  doi={10.1109/ISMAR62088.2024.00099}}

@inproceedings{jerald_human-centered_2017,
	title = {Human-centered design for immersive interactions},
	doi = {10.1109/VR.2017.7892361},
	booktitle = {2017 {IEEE} {Virtual} {Reality} ({VR})},
	author = {Jerald, Jason},
	year = {2017},
	pages = {431--432},
}



\end{document}